\documentclass[twoside, twocolumn, 10pt]{IEEEtran}

\usepackage[cmex10]{amsmath}
\usepackage{amssymb,amsthm,cite,graphicx,url, color,etoolbox}
\usepackage[tight,footnotesize]{subfigure}
\graphicspath{{./figures/}}
\theoremstyle{plain}
\newtheorem{lemma}{Lemma}
\newtheorem{theorem}{Theorem}

\newtheorem{proposition}{Proposition}

\theoremstyle{definition}
\newtheorem{definition}{Definition}
\theoremstyle{remark}
\newtheorem{remark}{Remark}

\title{A Constrained Random Demodulator for Sub-Nyquist Sampling}
\author{Andrew~Harms,~\IEEEmembership{Student Member,~IEEE}, Waheed~U.~Bajwa,~\IEEEmembership{Member,~IEEE}, and Robert~Calderbank,~\IEEEmembership{Fellow,~IEEE}%
\thanks{Copyright (c) 2012 IEEE. Personal use of this material is permitted. However, permission to use this material for any other purposes must be obtained from the IEEE by sending a request to pubs-permissions@ieee.org.}%
\thanks{This work was supported in part by ONR under Grant N00014-08-1-1110, by AFOSR under Grants FA9550-09-1-0551 and FA 9550-09-1-0422, by NSF under Grant DMS-0914892, and by DARPA under the KECoM Program.}%
\thanks{A.~Harms is with the Department of Electrical Engineering, Princeton University, Princeton, NJ 08544 USA (email: hharms@princeton.edu)}%
\thanks{W.~U.~Bajwa is with the Department of Electrical and Computer Engineering, Rutgers University, Piscataway, NJ 08854 USA (email: waheed.bajwa@rutgers.edu)}%
\thanks{R.~Calderbank is with the Department of Electrical and Computer Engineering, Duke University, Durham, NC 27708 USA (email: robert.calderbank@duke.edu).}}
\date{}

\begin{document}

\maketitle

\begin{abstract} 
This paper presents a significant modification to the Random Demodulator (RD) of Tropp et al. for sub-Nyquist sampling of frequency-sparse signals. The modification, termed constrained random demodulator, involves replacing the random waveform, essential to the operation of the RD, with a constrained random waveform that has limits on its switching rate because fast switching waveforms may be hard to generate cleanly.  The result is a relaxation on the hardware requirements with a slight, but manageable, decrease in the recovery guarantees. The paper also establishes the importance of properly choosing the statistics of the constrained random waveform.  If the power spectrum of the random waveform matches the distribution on the tones of the input signal (i.e., the distribution is proportional to the power spectrum), then recovery of the input signal tones is improved. The theoretical guarantees provided in the paper are validated through extensive numerical simulations and phase transition plots.
\end{abstract}


\IEEEpubid{0000--0000/00\$00.00~\copyright~2012 IEEE }

\IEEEpubidadjcol
\section{Introduction}
\IEEEPARstart{M}{odern} signal processing relies on the sampling of analog signals for discrete-time processing.  The standard approach to sampling signals is based on the Shannon--Nyquist sampling theorem, which states that a bandlimited signal can be faithfully reconstructed from its samples collected uniformly at the Nyquist rate.  However, this standard approach to sampling can be unwieldy for signals with very large bandwidths due to the physical constraints on modern Analog-to-Digital Converter (ADC) technology.  The rule of thumb in ADC technology is that a doubling of the sampling rate causes a 1 bit reduction in resolution \cite{le05} or, more explicitly, $P = 2^B\cdotp f_s$ where $B$ is the \emph{effective number of bits} (ENOB), a measure of resolution of an ADC, and $f_s$ is the sampling rate. This expression states that for a required sampling resolution, the sampling rate has a hard upper limit due to constraints on the ADC technology, and vice versa.  The constant $P$ is dependent on the particular ADC architecture and has steadily increased over time as the technology has improved; the current state-of-the-art allows for sampling at $1$ GHz with a resolution of approximately $10$ ENOB\cite{walden08,murmannADC}.  Unfortunately, this increase tends to happen rather slowly compared to the advancement seen in other areas of technology, such as microprocessor technology following Moore's law.  In particular, applications such as spectrum sensing for cognitive radios push modern ADC technology to its limit.

\IEEEpubidadjcol
\subsection{Random Demodulation for Sub-Nyquist Sampling}
\begin{figure}[!t]
  \centering
    \setlength{\unitlength}{.25in}
  \begin{picture}(12,4.8)
    \put(0,3){\makebox{$f(t)$}}
    \put(1.1,3.1){\vector(1,0){1.9}}
    \put(3,3){\makebox{$\bigotimes$}}
    \put(1.6,.75){\framebox{\small \begin{tabular}{c}Waveform \\ Generator\end{tabular} }}
    \put(3.3,1.6){\vector(0,1){1.25}}
    \put(1.7,2.0){\makebox{$p_m(t)$}}
    \put(2.2,3.8){\makebox{$f(t) \cdotp p_m(t)$}}
    \put(3.6,3.1){\vector(1,0){2.6}}
    \put(6.2,3){\framebox{$\int_{t-\frac{1}{R}}^{t}$}}
    \put(7.8,3.1){\vector(1,0){1.1}}
    \put(10,3.1){\line(-2,1){1}}
    \put(9.6,3.8){\vector(-1,-2){0.4}}
    \put(10,3.1){\vector(1,0){1.5}}
    \put(10,4){\makebox{$t = \frac{n}{R}$}}
    \put(11.7,3){\makebox{$y[n]$}}
  \end{picture}
  \caption{Block diagram of the random demodulator\cite{tropp10}: The input signal is multiplied by a waveform generated from a Rademacher chipping sequence, then low-pass filtered, and finally sampled at a sub-Nyquist rate $R \ll W$.}
  \label{fg:rd}

\end{figure}
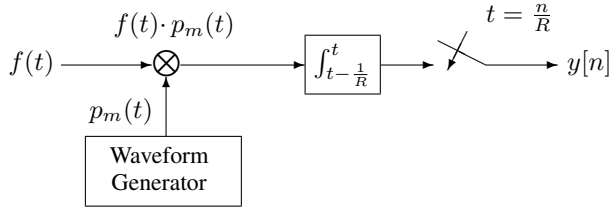

Though Nyquist sampling is the standard approach to sampling, other schemes have been considered that require a lower sampling rate for analog-to-digital conversion.  The key to the success of these schemes is leveraging additional prior information about the class of signals to be sampled (perhaps in addition to being bandlimited).  One such class of signals corresponds to complex-valued signals comprising a relatively small number of tones ($S$) in a very large (two-sided) bandwidth ($W$): $S \ll W$.  We say these signals have \emph{sparse} spectral content. This class of signals is of significant interest in applications such as spectrum sensing, and is the one we will concentrate on for the rest of this paper. We refer the reader to Section~\ref{sec:background} for a mathematically precise definition of this signal class.  Two good architectures to sample such signals are the Non-Uniform Sampler (NUS) \cite{yoo-candes12, candes-romberg-tao06} and the Random Demodulator (RD) \cite{tropp10}.  In this paper we concentrate exclusively on the RD because it offers a much more general framework for sub-Nyquist sampling.  The block diagram of the RD architecture is presented in Fig. \ref{fg:rd} and will be reviewed in more detail later\footnote{While we focus exclusively on a single-channel system, the analysis can be easily extended to the multi-channel setting.}. The major results for the RD can be summarized as follows \cite[Theorems 1 and 2]{tropp10}: let $\mathrm{C}$ be a positive, universal constant and let $W$ be the Nyquist rate.  The constituent tones of signals sampled by the RD can be recovered with high probability if the sampling rate $R$ scales as
($i$) $R\geq \mathrm{C}[S\log W + \log^3 W]$ for signals composed of $S$ randomly located tones\footnote{It is worth noting here that the NUS is shown to have similar results \cite[Theorem 1.3]{candes-romberg-tao06}.} and
($ii$) $R\geq \mathrm{C}S\log^6 W$ for signals composed of arbitrary $S$ tones.
Contrast these results to the Shannon--Nyquist sampling theorem, which guarantees recovery of the original signal from its samples if $R \geq W$.

A building block of the RD is a white noise-like, bipolar modulating waveform $p_m(t)$ (see Fig. \ref{fg:rd}).  This waveform switches polarity at the Nyquist rate of the input signal.  An implicit assumption is that this waveform, in the analog domain, is made up of perfect square pulses with amplitude either $+1$ or $-1$.  Hardware constraints, however, mean that a real waveform cannot switch polarity instantaneously and will encounter shape distortion.  A non-zero time $\tau$ is required to switch polarity and is dictated by the circuits encountered in ADC architectures\cite[Ch. 4]{martin2012}.  The transitions therefore occur over this time-scale, and the square waveform can be viewed as passing through a low-pass filter with a bandwidth proportional to $1/\tau$.  One implication is a reduction of the energy captured in the measurements that depends on $\tau$ and the number of transitions in the waveform.  For a larger $\tau$, or for more transitions in the waveform, less energy is captured in the measurements.

Over 30 years ago a similar problem affected the peak detection of binary signals written on magnetic media. In magnetic recording, data is recovered by passing a read head over the media; a higher recording density means there is greater interference between the read-back voltages of adjacent bits. To reduce distortion in the read back voltages, Tang and Bahl introduced \emph{Run-Length Limited} (RLL) sequences \cite{tang70}.  Run-length constraints specify the minimum separation $d$ and the maximum separation $k$ between transitions from one symbol to another (say $+1$ to $-1$).  Tang and Bahl proposed using these RLL sequences to increase the number of bits written on the magnetic medium by a factor of $d+1$ without affecting the read-back fidelity.  Note that RLL sequences, compared to unconstrained sequences, require a longer length to store the same amount of information.  Tang and Bahl nonetheless observed that for certain RLL sequences the fractional increase in length is smaller than $d + 1$, leading to a net increase in recording density because the allowed closer spacing of the physical bits (on the medium) overcomes the increase in bit-sequence length.  The reader may refer to \cite{immink98} for further details and a nice overview on this topic.

\subsection{Our Contributions: Constrained Random Demodulation}
\begin{figure}[!t]
  \centering
  \subfigure[An unconstrained sequence]{
    \includegraphics[width=2.5in]{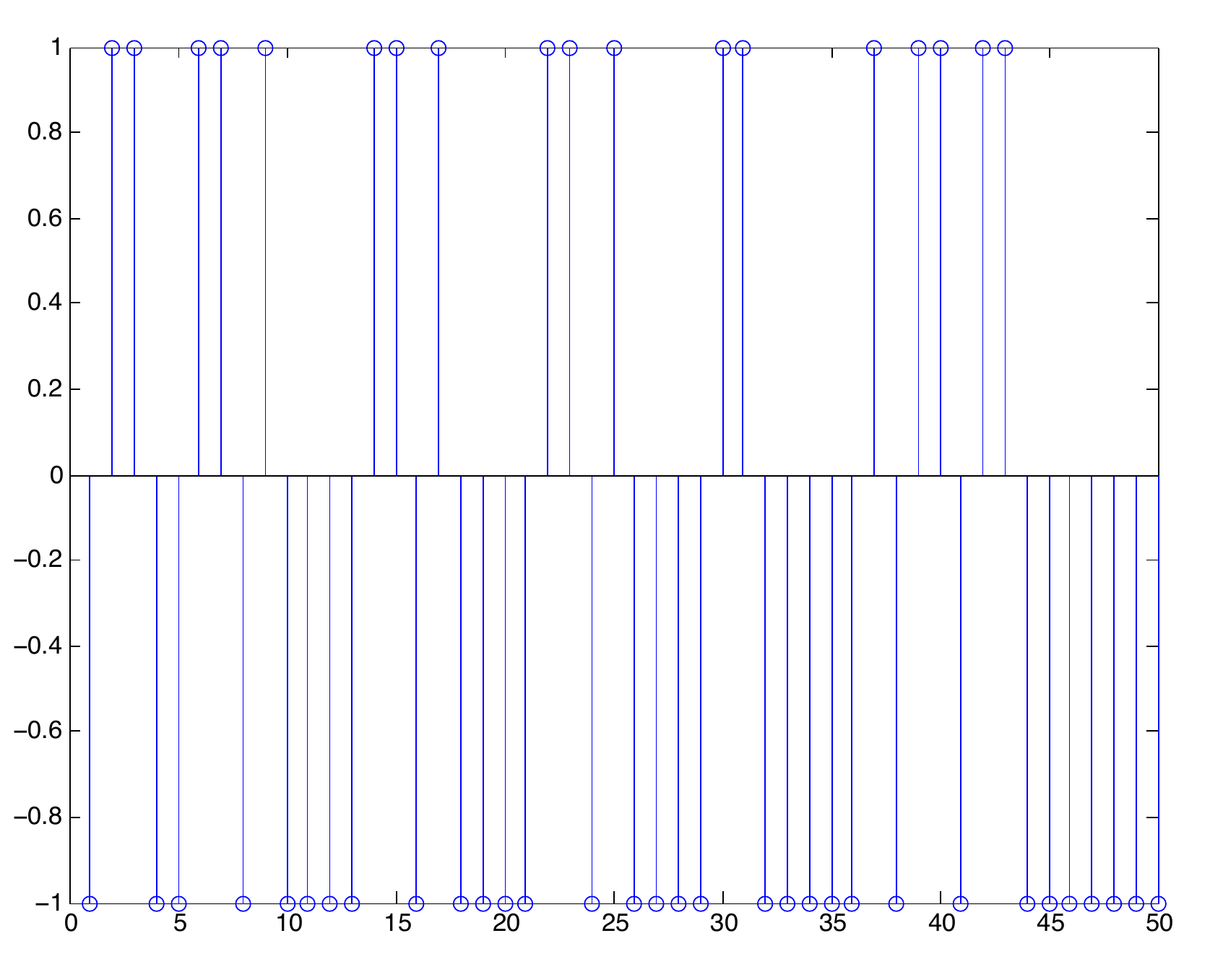}
    \label{fg:rd-waveform-example}
  }
  \subfigure[An RLL sequence with $(d,k)=(1,4)$]{
    \includegraphics[width=2.5in]{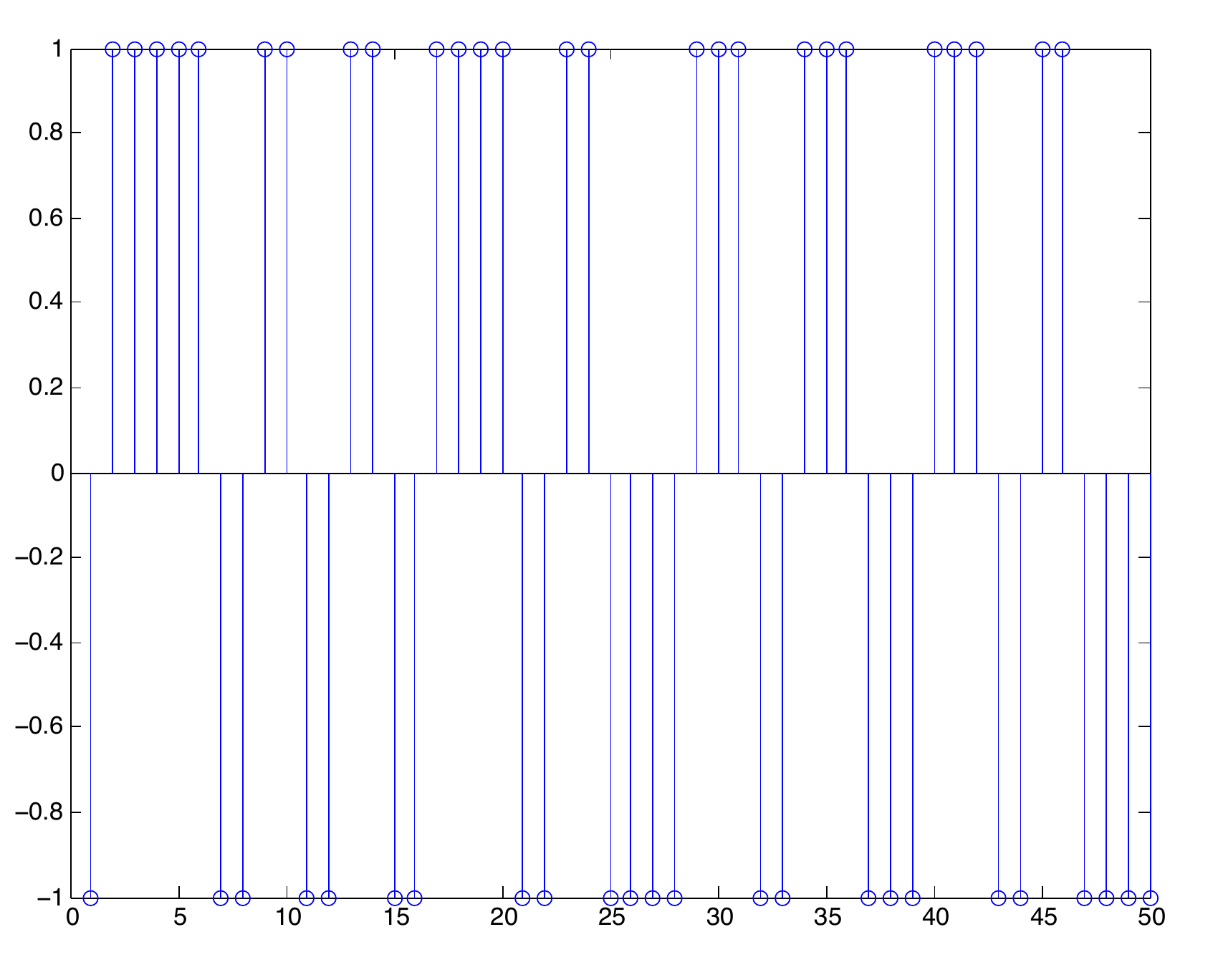}
    \label{fg:rll-waveform-example}
  }
  \caption{Comparing an RLL sequence to an unconstrained sequence: The unconstrained sequence can switch from one level, to the other, and back without limitation.  The RLL sequence, on the other hand, remains at a level for at least $d+1$, and at most $k+1$, time instants after it changes levels and cannot switch back right away.  The (analog) modulating waveform is generated using a shifted square pulse with the appropriate sign from the sequence.}
  \label{fg:rd-vs-rll-waveform}
\end{figure}

In this paper, we make two major contributions to the area of sub-Nyquist sampling for signals with sparse spectral content.  Our first contribution is to apply the lessons learned from magnetic recording to the RD.  Specifically, we replace the modulating waveform of the RD with a $(d,k)$-constrained waveform generated from an RLL sequence (see Fig. \ref{fg:rd-vs-rll-waveform}).  We refer to such a sampling system as a \emph{Constrained Random Demodulator} (CRD).  The use of an RLL sequence reduces the average number of transitions in the waveform by a factor of $d+1$, which results in an increase in the signal energy captured by the hardware.  From another viewpoint, if we fix the acceptable energy loss (or average number of transitions in the waveform), then using an RLL sequence allows a larger input signal bandwidth.  We do, of course, pay a price: an RLL sequence introduces statistical dependence across the waveform.  Our first major contribution is therefore establishing that the CRD still enjoys some theoretical guarantees for certain choices of waveform.  In fact, we explicitly show that the power spectrum of the waveform is the key to understanding these guarantees and, hence, to choosing the best RLL sequence.  Further, we outline a tradeoff in acquirable bandwidth versus sparsity of the input signal and show through numerical simulations that a $20\%$ increase in the bandwidth can be handled by the CRD with a negligible decrease in average performance.  Our work here builds upon our preliminary work in \cite{harms11-icassp, harms-camsap11} that was primarily limited to introducing the idea of the CRD along with Theorem \ref{thm:ripcord} (without proof).

\begin{remark}
Heuristically, the theoretical guarantees in this paper rely on two things: (i) each (active) tone leaves an identifiable signature that can be extracted from the measurements and (ii) the measurements capture a significant amount of energy of each tone.  We will show that the identifiability depends on the modulating sequence power spectrum.  Once this is established, we would further like to maximize the captured energy.  Since an RLL waveform leads to an increase in the captured energy because of the switching constraints previously discussed, its use in a hardware implementation will lead to improved performance as long as it satisfies the identifiability criterion.
\end{remark}

Our second contribution is laying down the foundations of a concept that we call \emph{Knowledge-Enhanced Compressive Sensing} (KECoM) for sub-Nyquist sampling, which we preliminarily explored in \cite{harms-camsap11} with limited numerical experiments.  In the context of the CRD, the principle of KECoM assumes that some tones in the input signal are statistically more likely to appear than others.  An immediate application of this is a spectrum sensing problem where some regions of the spectrum are assigned a higher detection priority than others, but none are deemed uninformative.  We show through numerical simulations that the distribution of the tones in the input signal has a profound effect on the reconstruction of input signals from samples collected using a CRD.  Specifically, we show with phase transition plots \cite{donoho09} that if the prior distribution over the tones matches the power spectrum of the RLL sequence used by the CRD, then the reconstruction performance improves when compared to a uniform distribution over the tones. Note that \cite{setti10, setti11} have also recently explored ideas along similar lines, albeit for a different class of sequences. In contrast to \cite{setti10, setti11}, we provide a theoretical analysis and, additionally, a comprehensive numerical analysis of RLL sequences in the RD by examining the phase transition plots.

\subsection{Other Sub-Nyquist Sampling Schemes}
The work of Rife and Boorstyn \cite{rife74} in the mid-70's is an early example of a successful sub-Nyquist sampling scheme.  Their goal was to take samples of a sinusoid at a sub-Nyquist rate and then perform parameter estimation to determine the amplitude, frequency, and phase of a single, unknown tone. They also extended their work to the case of multiple tones in a large bandwidth \cite{rife76}.  Their work, however, becomes intractable when considering more than a couple tones.  This is an early example of what has become known as compressed sensing of sparse signals.  \emph{Compressed Sensing} (CS) is the systematic exploration of sparsity as a prior model for input signals and recovery of these signals from a small number of linear measurements \cite{duarte11}.  It has produced many analytical tools and algorithms for signal recovery.  In addition to the RD, several other sub-Nyquist sampling architectures have taken advantage of ideas from CS including \emph{Chirp Sampling} \cite{applebaum08} and \emph{Xampling} \cite{eldar09}.

While the RD considers a bandlimited input signal model with few active tones, several other classes of signals have been considered in the literature with the goal of finding more efficient sampling methods.  One such class contains signals with so-called ``Finite Rates of Innovation" \cite{vetterli02}.  Signals belonging to this class can be described by a finite number of degrees of freedom in a given time interval, and it has been shown that they can be reconstructed from a small number of samples that is proportional to the degrees of freedom in that time interval.  Another class constitutes signals in ``shift-invariant subspaces."  These signals are composed of a superposition of shifted `generator' functions (e.g., splines or wavelets); see \cite{unser2000} for a nice overview of this signal class.  In \cite{blu-unser-1999} and \cite{blu-unser-part2-1999}, this signal model is shown to provide an alternative to the bandlimited signal model; in particular, it allows the reconstruction of signals belonging to Sobolev spaces with an approximation error that scales polynomially with the sampling period.

One possible drawback to utilizing the RD for sampling is its assumed discrete-frequency signal model (cf.~Section~\ref{sec:background}).  Specifically, the RD assumes that the input signal can be described by a discrete set of integral frequencies, while real-world signals are likely to contain tones off this grid. While this signal model might not entirely describe real-world signals, the effectiveness of the RD architecture has been successfully demonstrated in the lab \cite{nichols11,jyoo-RMPI}.  To address signals with tones not conformant to the integral-frequency assumption, we consider energy leakage in the frequency domain.  A tone that does not fall exactly on the assumed frequency grid will leak energy across several tones due to the inherent windowing. The result is that a signal which is $S$-sparse in the analog domain becomes $(aS)$-sparse after being sampled, where $a > 1$. Other schemes, such as Xampling \cite{eldar09}, offer an alternative approach assuming a different signal model; the pros and cons of both systems are examined in \cite{lexa11}. While our focus in this paper is exclusively on the RD, we believe that our contributions could have implications for other sub-Nyquist architectures.  Specifically, the Xampling architecture uses modulating sequences similar to the ones used in the RD/CRD, and we believe that RLL sequences could benefit the Xampling architecture as well.  A detailed analysis is, however, beyond the scope of this paper.

We would also like to point to a possible utility of RLL sequences in the NUS.  The implementation described in \cite{yoo-candes12} requires a minimum and maximum spacing between sample points while the analysis in \cite{candes-romberg-tao06} assumes the sample points are uniformly random without any constraints.  The constraints in \cite{yoo-candes12} can thus be described by an RLL sequence made up of 0's and 1's with 1's representing sampling points.  We feel this interpretation of the limitations in [4] can help us mathematically analyze the architecture in [4], but a detailed investigation of this is beyond the scope of this paper.

\subsection{Organization and Notation}
The remainder of the paper is organized as follows.  We first provide some background on the RD in Section \ref{sec:background} and then explain the challenges encountered by introducing RLL sequences into the RD architecture in Section \ref{sec:CRD}.  We then present our main theoretical results in Section \ref{sec:CRD} and two examples of constrained sequences, one with bad results (Section \ref{sec:RCS}) and one with good results (Section \ref{sec:general-sequences}), to illustrate the effectiveness of our analysis. Finally, in Sections \ref{sec:numerical-results} and \ref{sec:kecom-sampling} we present numerical simulations to offer some verification of the theoretical results.

In the following we denote matrices with upper case roman letters and vectors with lower case roman letters.  Scalars are denoted with italic lower case letters.  We write $^*$ for the conjugate transpose of a matrix, vector, or scalar.  We reserve the letters $\mathrm{C}$ and $\mathrm{c}$ in roman font to denote universal constants that could change values at each instance.  For a matrix, $\mathrm{A}|_{\Omega\times\Omega}$ denotes the principal submatrix of $\mathrm{A}$ created from the columns/rows given in $\Omega$.  We also use $||\cdot||$ for the spectral norm of a matrix and $||\cdot||_{\max}$ for the maximum absolute entry of a matrix.  For a random variable $\mathrm{B}$, let $\mathbb{E}[\mathrm{B}]$ be the expectation and $\mathbb{E}^p\mathrm{B} = (\mathbb{E}|\mathrm{B}|^p)^{1/p}$.  Let $\mathbb{P}\{\cdot\}$ denote the probability of an event.  The short-hand $j\sim r$ means $(r-1)W/R < j \leq rW/R$ for some $W$ and $R$ such that $R$ divides $W$.

\section{Background: The Random Demodulator}\label{sec:background}
We start with a brief review of the RD architecture and highlight the key components that allow sampling of sparse, bandlimited signals and refer the reader to \cite{tropp10} for a thorough overview.  To start, the RD takes samples at a sub-Nyquist rate $R$ while retaining the ability to reconstruct signals that are periodic, (two-sided) bandlimited to $W$ Hz, and completely described by a total of $S \ll W$ tones.  These conditions describe a large class of wide-band analog signals comprised of frequencies that are small in number relative to the total bandwidth but are at unknown locations.

Formally, the input signal to a RD takes the following parametric form
  \begin{equation}\label{eq:sig_model}
    f(t)=\sum_{\omega \in \Omega}a_{\omega}e^{-2\pi \imath \omega t},\ t \in [0,1)
  \end{equation}
where $\Omega \subset \{0,\pm1,...,\pm W/2-1, W/2\}$\footnote{We assume $W$ is even.  An appropriate change of the set $\Omega$ would cover the case of $W$ odd.} is a set of $S$ integer-valued frequencies and $\{a_{\omega} : \omega \in \Omega\}$ is a set of complex-valued amplitudes.  Fig. \ref{fg:rd} illustrates the actions performed by the RD.  The input $f(t)$ is first multiplied by
$$p_m(t) = \sum_{n=0}^{W-1}\varepsilon_n 1_{\left[\frac{n}{W},\frac{n+1}{W}\right)}(t),$$
where the discrete-time \emph{modulating sequence} $\varepsilon = [\varepsilon_n]$ is a Rademacher sequence, a random sequence of independent entries taking values $\pm 1$ equally likely.  Next, the continuous-time product $f(t)\cdotp p_m(t)$ is low-pass filtered using an ``integrate and dump'' filter.\footnote{It can be easily shown that the frequency response of this filter tapers off at high frequencies.  Hence, it is a low-pass filter.}  Finally, samples are taken at the output of the low-pass filter at a rate of $R \ll W$ to obtain $y[n]$.

\subsection{Matrix Representation of the Random Demodulator}\label{sec:mat_rep}
One of the major contributions of \cite{tropp10} is expressing the actions of the RD on a continuous-time, sparse, and bandlimited signal $f(t)$ in terms of the actions of an $R \times W$ matrix $\mathrm{\Phi_{RD}}$ on a vector $\mathrm{\alpha} \in \mathbb{C}^W$ that has only $S$ nonzero entries. Specifically, let $\mathrm{x} \in \mathbb{C}^W$ denote a Nyquist-sampled version of the continuous-time input signal $f(t)$ so that $\mathrm{x}_n = f(\frac{n}{W})$, $n = 0,\cdots,W-1$. It is then easy to conclude from \eqref{eq:sig_model} that $\mathrm{x}$ can be written as $\mathrm{x} = \mathrm{F} \alpha$, where the matrix
$$\mathrm{F} = \frac{1}{\sqrt{W}} \left[e^{-2\pi \imath n\omega/W}\right]_{(n,\omega)}$$
denotes a (unitary) discrete Fourier transform matrix and $\alpha \in \mathbb{C}^W$ has only $S$ nonzero entries corresponding to the amplitudes, $a_{\omega}$, of the nonzero frequencies in $f(t)$. Next, the effect of multiplying $f(t)$ with $p_m(t)$ in continuous-time is equivalent in the discrete-time Shannon--Nyquist world to multiplying a $W \times W$ diagonal matrix $\mathrm{D} = \text{diag}(\varepsilon_0, \varepsilon_1, \cdots, \varepsilon_{W-1})$ with $\mathrm{x} = \mathrm{F} \alpha$. Finally, the effect of the integrating filter on $f(t) \cdot p_m(t)$ in the discrete-time Shannon--Nyquist setup is equivalent to multiplying an $R \times W$ matrix $\mathrm{H}$, which has $W/R$ consecutive ones starting at position $rW/R+1$ in the $r^{th}$ row of $\mathrm{H}$ and zeros elsewhere, with $\mathrm{D}\mathrm{F}\alpha$.\footnote{Throughout this paper we assume that $R$ divides $W$; otherwise, a slight modification can be made to $\mathrm{H}$ as discussed in \cite{tropp10}.} An example of $\mathrm{H}$ for $R=3$ and $W=9$ is
  $$\mathrm{H}=\begin{bmatrix}1&1&1&&&&&& \\
  				&&&1&1&1&&& \\
				&&&&&&1&1&1
	\end{bmatrix}
  $$
The RD collects $R$ samples per second, and therefore, the $R$ samples collected over 1 second at the output of the RD can be collected into a vector $\mathrm{y} \in \mathbb{C}^R$.  It follows from the preceding discussion that $\mathrm{y} = \mathrm{H}\mathrm{D}\mathrm{F}\alpha = \mathrm{\Phi_{RD}}\cdot\alpha$, where we have the complex-valued \emph{random demodulator matrix} $\mathrm{\Phi_{RD}} = \mathrm{H}\mathrm{D}\mathrm{F}$.

\subsection{Signal Recovery}\label{sec:sig_rec}
Given the discrete-time representation $\mathrm{y} = \mathrm{\Phi_{RD}}\cdotp\alpha$, recovering the continuous-time signal $f(t)$ described in \eqref{eq:sig_model} is equivalent to recovering the $S$-sparse vector $\alpha$ from $\mathrm{y}$. In this regard, the primary objective of the RD is to guarantee that $\alpha$ can be recovered from $\mathrm{y}$ even when the sampling rate $R$ is far below the Nyquist rate $W$.  Recent theoretical developments in the area of CS provide us with greedy as well as convex optimization-based methods that are guaranteed to recover $\alpha$ (or a good approximation of $\alpha$) from $\mathrm{y}$ (possibly in the presence of noise) as long as the \emph{sensing matrix} $\mathrm{\Phi_{RD}}$ satisfies certain geometrical properties \cite{duarte11}.  Tropp et al. \cite{tropp10} uses two properties from the CS literature to analyze the sensing matrix.  The first is the \emph{coherence}.  The coherence $\mu$ of a matrix $\mathrm{\Phi}$ is the largest inner product between its (scaled to unit-norm) columns $\phi_{\omega}$: $\mu = \max_{\omega\neq\alpha}|\langle\phi_{\omega},\phi_{\alpha}\rangle |$.  Many recovery algorithms rely on the coherence of the sensing matrix being sufficiently small \cite{tropp08-condofrandict}.  The analysis in \cite{tropp10} in this regard also relies on the input signals conforming to a \emph{random signal model}: given the signal model \eqref{eq:sig_model}, the index $\Omega$ is a set of $S$ tones drawn uniformly at random from the set of $W$ possible tones.  Further, the coefficients $a_{\omega}$ are drawn uniformly at random from the complex unit circle.  Under this signal model, $S$-sparse signals are recoverable with high probability if the sampling rate scales as $R \geq \mathrm{C}[S\log W + \log^3 W]$\cite{tropp10}.

The second property used in \cite{tropp10} is the \emph{Restricted Isometry Property} (RIP)\cite{candes05}.
\begin{definition}
The RIP of order $S$ with restricted isometry constant $\delta_S \in (0,1)$ is satisfied for a matrix $\Phi$ with unit-norm columns if $$(1-\delta_S)||\mathrm{x}||_2^2 \leq ||\mathrm{\Phi}\mathrm{x}||_2^2 \leq (1+\delta_S)||\mathrm{x}||_2^2$$ or equivalently
  \begin{equation}\label{eq:RIP_def}
    \left \lvert \frac{\lVert\Phi \mathrm{x}\rVert_2^2 - \lVert\mathrm{x}\rVert_2^2}{\lVert\mathrm{x}\rVert_2^2}\right \rvert \leq \delta_S
  \end{equation}
for every $\mathrm{x}$ with $\lVert\mathrm{x}\rVert_0 \leq S$.  Here, $||\mathrm{x}||_0$ counts the number of non-zero entries in $\mathrm{x}$.
\end{definition}
Note that RIP-based analysis tends to be stronger than the coherence-based analysis because the RIP provides a better handle on worst-case performance as well as on performance in the presence of noise\cite[Theorem 1]{candes-StableRecovery}.  It also provides stable recovery even if the signal is not exactly sparse, but is well-described by a sparse signal (so-called compressible signals)\cite[Theorem 2]{candes-StableRecovery}.  We will therefore focus only on proving the RIP with the understanding that RIP automatically implies stable and robust recovery (see \cite{duarte11} and the references therein for a more extensive list of results).

In this paper, we use the ``triple-bar" norm of \cite{tropp10} to describe the RIP condition.  Given a matrix $\mathrm{A}$ and set of indices $\Omega \subset \{0,\ldots, W-1\}$, the triple-bar norm captures the least upper bound on the spectral norm of any $S\times S$ principal submatrix of $\mathrm{A}$:
\begin{equation}\label{eq:triplebarnorm}
  |||\mathrm{A}||| = \sup_{|\Omega| \leq S}\lVert \mathrm{A}|_{\Omega\times\Omega}\rVert.
\end{equation}
It can easily be checked that $|||\cdot|||$ is a norm and that (\ref{eq:RIP_def}) is satisfied if and only if $|||\Phi^*\Phi - \mathrm{I}||| \leq \delta_S$.

The main result of \cite{tropp10} in this respect is that the RD matrix satisfies the RIP of order $S$ as long as the sampling rate $R$ scales as $R \geq \mathrm{C}S\log^6W$.

\section{Constrained Random Demodulator}\label{sec:CRD}
As described in the previous section, the RD uses a random waveform generated from a Rademacher sequence with transition density of $\frac{1}{2}$ (on average, one transition every 2 Nyquist periods).  However, limitations of analog circuits imply that each transition in the waveform results in a loss of energy compared to a waveform with ideal square pulses\cite{martin2012}.  RLL sequences are an attractive way to generate waveforms with a reduced transition density of $\frac{1}{d+2}$.  Additionally, we will later show that RLL sequences can also lead to superior performance for specific classes of input signals.  We remind the reader that if an RLL sequence is used we call the resulting system a \emph{Constrained Random Demodulator} (CRD) and denote the corresponding system matrix as $\mathrm{\Phi_{CRD}} = \mathrm{HDF}$ where $\mathrm{D}$ contains an RLL sequence $\varepsilon$ instead of a Rademacher sequence.  The properties of the Rademacher sequence, in particular independence, are central to the analysis of the RD in \cite{tropp10}; we therefore must carefully consider the impact of using an RLL sequence that is inherently correlated.

The strength of \cite{tropp10} is that it shows that the RD matrix satisfies the RIP with high probability, allowing strong guarantees to be made about the recovery of signals sampled with the RD.  The RIP is satisfied primarily because of three properties of the RD matrix:
($i$) the Gram matrix averages (over realizations of the modulating sequence) to the identity matrix,
($ii$) the rows are statistically independent, and
($iii$) the entries are uniformly bounded.
All three properties rely on the independence of the modulating sequence.  In the CRD, we have to deal with dependence across $\varepsilon$.  Nevertheless, the last two properties are handled relatively easily.  Specifically, if we can find some distance between entries in $\varepsilon$ such that any two entries, when separated by this distance, are independent, then we can partition the rows of $\mathrm{\Phi_{CRD}}$ (or entries of $\varepsilon$) into sets of independent rows (entries).  We can then find bounds similar to those found in \cite{tropp10} for these sets and take a union bound over all the sets to obtain the desired properties.

\subsection{Maximum Dependence Distance}
To make the previous discussion more concrete, recall that the $(r,\omega)$ entry of $\mathrm{\Phi_{CRD}}$ is
\begin{equation}\label{eq:phi-entry}
  \varphi_{r\omega} = \sum_{j\sim r}\varepsilon_jf_{j\omega}.
\end{equation}
If $\varepsilon$ is an independent sequence, then each $\varphi_{r\omega}$ is a sum of independent random variables, and each row of $\mathrm{\Phi_{CRD}}$ is independent.  However, if we use a correlated sequence then the rows may not be independent, and it is important to know the extent of the dependence within the sequence.
\begin{definition}
The \emph{Maximum Dependence Distance} (MDD), $\ell$, for a modulating sequence $\varepsilon$ is the smallest $\ell$ such that $\mathbb{E}[\varepsilon_j\varepsilon_{j+k}] = 0$ for all $j$ and $|k| \geq \ell$ \footnote{Note that this is a correlation distance, but that for the bipolar sequences of our concern, uncorrelated implies independent.}.
\end{definition}
Now, if we define $\rho = \lceil \frac{R}{W}(\ell-1)\rceil \leq (\ell-1)$, then any two rows of $\mathrm{\Phi_{CRD}}$ separated by at least $\rho+1$ rows will be independent.  Given $\rho$ and $\ell$, we can now partition the rows of $\mathrm{\Phi_{CRD}}$ into $\rho+1$ subsets where the rows in each subset are independent.\footnote{We assume for convenience that $\rho+1$ divides $R$.  This can be readily relaxed by adjusting the size of the last subset.}  Using this partitioning scheme, we can proceed with the analysis of independent rows and finally take a union bound over all subsets.  Using $\ell$, we can similarly show that each entry of $\mathrm{\Phi_{CRD}}$ is uniformly bounded.  The details are in Appendices \ref{app:rip-proof} and \ref{app:random-signal-model}.

\subsection{The Gram Matrix}
Analysis of the Gram matrix of $\mathrm{\Phi_{CRD}}$ is a little more involved.  To start, denote the columns of $\mathrm{\Phi_{CRD}}$ by $\mathrm{\phi}_{\omega}$ and note that the $(r,\omega)$ entry of $\mathrm{\Phi_{CRD}}$ is given by \eqref{eq:phi-entry}.   The Gram matrix is a tabulation of the inner products between the columns and (as calculated in\cite{tropp10}) is given by $\mathrm{\Phi^*_{CRD}}\mathrm{\Phi_{CRD}} = \mathrm{I} + \mathrm{X}$.  Here, the $(\alpha,\omega)$ entry of $\mathrm{X}$ is the sum
\begin{equation}\label{eq:matrixX}
  x_{\alpha\omega} = \sum_{j\neq k}\varepsilon_j\varepsilon_k\eta_{jk}f^*_{j\alpha}f_{k\omega}
\end{equation}
where $[\varepsilon_0,\cdots,\varepsilon_{W-1}] = \varepsilon$ is the modulating sequence, $\eta_{jk} = \langle h_j,h_k\rangle$ with $h_j$ being the $j$th column of $\mathrm{H}$, and $f_{j\alpha}$ is the $(j,\alpha)$ entry of the (unitary) Fourier matrix $\mathrm{F}$.  Expanding $\eta_{jk}$, we have that
\begin{equation}\label{eq:windowing-function}
  \eta_{jk} = \begin{cases} 1, &  \frac{W}{R}r \leq j,k < \frac{W}{R}(r+1) \\ 0, & \text{otherwise} \end{cases}
\end{equation}
for each $r = 0,\cdots,R-1$.  We see that $\eta_{jk}$ acts as a `windowing' function in the sum.  In expectation, the Gram matrix is $\mathbb{E}[\mathrm{\Phi_{CRD}^*}\mathrm{\Phi_{CRD}}] = \mathrm{I} + \mathbb{E}[\mathrm{X}] = \mathrm{I} + \Delta$ where we have identified $\Delta \equiv \mathbb{E}[\mathrm{X}]$ with entries
\begin{equation}\label{eq:delta_def}
  \Delta_{\alpha\omega} = \sum_{j\neq k}\eta_{jk}f^*_{j\alpha}f_{k\omega}\mathbb{E}[\varepsilon_j\varepsilon_k].
\end{equation}

Note that $\Delta$ is completely determined by the autocorrelation of $\varepsilon$.  If an independent $\varepsilon$ is used (such as for the RD) then $\mathbb{E}[\varepsilon_j\varepsilon_k] = 0$ for $j\neq k$, $\Delta = \mathrm{0}$, and $\mathbb{E}[\mathrm{\Phi_{RD}^*}\mathrm{\Phi_{RD}}] = \mathrm{I}$.  In \cite{tropp10}, this relation is taken to mean that the columns of $\mathrm{\Phi_{RD}}$ form an orthonormal system in expectation. This can of course never be true if $R<W$, and the RIP is shown by bounding the deviation from this expectation in $|||\cdot|||$.

If $\varepsilon$ has non-zero correlations, however, then $\Delta$ does not disappear and the expectation of the Gram matrix is not the identity matrix.  To establish the RIP in this case, we still need to bound the deviation of the Gram matrix from the identity matrix, but now we must also contend with $\Delta$.  Nevertheless, if this matrix is small in $|||\cdot|||$ then our task is easier.  Since the autocorrelation of $\varepsilon$ determines $\Delta$, we want to choose a $\varepsilon$ that produces small $|||\Delta|||$.  In particular, recall that the RIP of order $S$ is satisfied if
\begin{equation}\label{eq:rip-delta}
  |||\mathrm{\Phi^*_{CRD}}\mathrm{\Phi_{CRD}} - \mathrm{I}||| \leq \delta_S.
\end{equation}
Expressing $\mathrm{I} = \mathbb{E}[\mathrm{\Phi_{CRD}^*\Phi_{CRD}}] - \Delta$, the left-hand side of \eqref{eq:rip-delta} can be bounded as
\begin{align}
  |||&\mathrm{\Phi^*_{CRD}}\mathrm{\Phi_{CRD}} - \mathrm{I}||| \notag \\
  &= |||\mathrm{\Phi^*_{CRD}}\mathrm{\Phi_{CRD}} - \mathbb{E}[\mathrm{\Phi^*_{CRD}}\mathrm{\Phi_{CRD}}] + \mathrm{\Delta}||| \notag \\
  &\leq |||\mathrm{\Phi^*_{CRD}}\mathrm{\Phi_{CRD}} - \mathbb{E}[\mathrm{\Phi^*_{CRD}}\mathrm{\Phi_{CRD}}]||| + |||\mathrm{\Delta}||| \label{eq:RIP-with-Delta}
\end{align}
due to the triangle inequality.  Therefore, to show the RIP we must upper bound the two terms in \eqref{eq:RIP-with-Delta}.  The first term will be bounded using an argument very similar to that used in \cite{tropp10} but modified to deal with the correlations in $\varepsilon$. Since the second term, $|||\Delta|||$, is determined by the autocorrelation of $\varepsilon$, we will provide a bound on $|||\Delta|||$ that directly relates to the choice of $\varepsilon$.

\subsection{Main Results}\label{sec:main-results}
The preceding discussion on $\ell$ and $\Delta$ enables us to make a statement about the RIP of a CRD that uses a correlated modulating sequence.

\begin{theorem}[RIP for the CRD] \label{thm:ripcord}
Let $\mathrm{\Phi_{CRD}}$ be an $R \times W$ CRD matrix using a modulating sequence with maximum dependence distance $\ell$ and $\Delta$ (as defined by \eqref{eq:delta_def}).  Next, pick $\delta, \delta^{\prime} \in (0,1)$ such that $\delta' < \delta - |||\Delta|||$ and suppose that $R$ divides $W$, $\ell$ divides $\frac{W}{R}$,\footnote{Throughout this paper, these requirements can be readily relaxed through meticulous accounting of terms in the analysis.} and $R$ satisfies
\begin{equation}\label{eq:thoerem1_R}
  R \geq \ell^3\delta'^{-2}\cdotp \mathrm{C}\cdotp S\log^6(W)
\end{equation}
where $\mathrm{C}$ is a positive constant.  Then with probability $1-\mathcal{O}(W^{-1})$ the CRD matrix $\mathrm{\Phi_{CRD}}$ satisfies the RIP of order $S$ with constant $\delta_S\leq\delta$.
\end{theorem}

The proof is provided in Appendix \ref{app:rip-proof}.  As with the RD, the sampling rate $R$ must scale linearly with the sparsity $S$ of the input signal and (poly)logarithmically with the bandwidth $W$.  The sampling rate, however, also depends on the maximum correlation distance $\ell$ and on the matrix $\Delta$.  Both of these are determined by the choice of $\varepsilon$.  If we choose an independent (i.e., unconstrained) $\varepsilon$, then $\ell = 1$, $\Delta = 0$ and we get back the RD result of \cite{tropp10}. For a constrained $\varepsilon$, we must restrict ourselves to sequences such that $\Delta$ satisfies $|||\Delta||| < 1$.  Obviously we would like to find sequences for which both $\ell$ and $|||\Delta|||$ are as small as possible. With this criterion in mind, in the next two sections we will examine two classes of sequences to see how well they work in the CRD framework.

In addition to the RIP, we also use the coherence of the sensing matrix to provide results for the random signal model described in Section \ref{sec:background}.  In the sequel, we use a matrix from \cite{samson00} to capture the dependence in $\varepsilon$.  For a sequence $\varepsilon$, define the triangular matrix $\Gamma$ of ``mixing coefficients" as $\Gamma = \{\gamma_{ij}\}$ with
\begin{equation*}
  \gamma_{ij} = \begin{cases} 0, & i>j \\
  					       1, & i=j \\
					       \big|\mathbb{P}\left(\varepsilon_{j} = +1|\varepsilon_i=-1\right) \\ - \mathbb{P}\left(\varepsilon_{j} = +1|\varepsilon_i=+1\right)\big|, & i<j.
  \end{cases}
\end{equation*}

\begin{theorem}[Recovery under the random signal model]
\label{thm:rand-sig-coherence}
Suppose that the sampling rate satisfies
\begin{equation}
  R \geq \mathrm{C}\ell^2[S\log W + \log^3 W]
\end{equation}
for some positive constant $\mathrm{C}$, and that $R$ divides $W$ and $\ell$ divides $\frac{W}{R}$.  Also suppose that $W$ satisfies
\begin{equation}\label{eq:W-constraint-gamma}
  \frac{\log^2 W}{\sqrt{W}} \leq \frac{\mathrm{C}}{16\sqrt{(\ell-1)}||\Gamma||^2}.
\end{equation}
Now, let $\mathrm{\alpha}$ be a vector with $S$ non-zero components drawn according to the random signal model in Section \ref{sec:sig_rec}, and let $\mathrm{\Phi_{CRD}}$ be an $R\times W$ CRD matrix using a stationary modulating sequence with maximum dependence distance $\ell$.  Let $\mathrm{y} = \mathrm{\Phi_{CRD}} \cdotp\mathrm{\alpha}$ be the samples collected by the CRD.  The solution to the convex program
\begin{equation}\label{eq:convex_program}
  \hat{\mathrm{\alpha}} = \arg\min_{\mathrm{v}} ||\mathrm{v}||_1 \ \text{subject to} \ \mathrm{\Phi_{CRD}}\mathrm{v} = \mathrm{y}
\end{equation}
satisfies $\hat{\mathrm{\alpha}} = \mathrm{\alpha}$ with probability $1-\mathcal{O}(W^{-1})$.
\end{theorem}
The proof is given in Appendix \ref{app:random-signal-model}.  The bounds offered here are similar to those in \cite{tropp10} with the rate scaling linearly with the sparsity $S$ and logarithmically with the bandwidth $W$ but more tightly constrained by the factor of $\ell^2$ and the extra constraint on $W$.

Because the choice of modulating sequence plays such a pivotal role in our analysis of the CRD, a natural question is what types of sequences offer good performance and what types offer bad performance.  In the sequel, we analyze two different types of sequences: one for which Theorems \ref{thm:ripcord} and \ref{thm:rand-sig-coherence} (approximately) apply, and one for which they do not.  Numerical experiments in Section \ref{sec:numerical-results} then show that these results appear to be tight.  Nevertheless, we must stress two points here.  First, Theorems \ref{thm:ripcord} and \ref{thm:rand-sig-coherence} are only sufficient conditions on the sampling rate and modulating sequence; a different analysis could offer stronger results.  Second, the modulating sequences that are shown to work well numerically satisfy Theorems \ref{thm:ripcord} and \ref{thm:rand-sig-coherence} in an approximate sense.  From an engineering perspective, however, the approximation (discussed in Sec. \ref{sec:general-sequences}) is well justified and validated further by the numerical experiments.

\section{Repetition-coded Sequences}\label{sec:RCS}
We begin by analyzing sequences that satisfy the RLL constraints and have a small value of $\ell$ but have a large $|||\Delta|||$ and do not satisfy Theorems \ref{thm:ripcord} or \ref{thm:rand-sig-coherence}.
\begin{definition}
A \emph{repetition-coded sequence} (RCS) is generated from a Rademacher sequence by repeating each element $d$ times.  Let the repetition-coded sequence be denoted as $\mathrm{\varepsilon_{RCS}} = [\varepsilon_0,\ldots,\varepsilon_{W-1}]$ and let $[\varepsilon_{(d+1)n}]$, $0\leq n \leq\frac{W}{d+1}-1$ be a Rademacher sequence.  We then require for $1\leq i \leq d$ and each $n$ that
\begin{equation}\label{eq:rcs-definition}
  \varepsilon_{(d+1)n} = \varepsilon_{(d+1)n+i}.
\end{equation}

\end{definition}
Such a sequence switches at a rate of $W/(d+1)$.  We discuss these sequences because they are one of the simplest forms of RLL sequences and also have very small MDD.  To see this, notice that each group of repeated elements, $[\varepsilon_{(d+1)n+i}]$ for $0\leq i \leq d$, is completely dependent while independent of every other element in the sequence.  The maximum dependence distance is $\ell = d+1$.

Since the performance of the CRD also depends upon $|||\Delta|||$, we need to bound $|||\Delta|||$ and understand its behavior.  To start, assume that $R$ divides $W$ and $\ell$ divides $\frac{W}{R}$ and denote by $\mathrm{\varepsilon_{RCS}}$ an RCS.  Let $\mathrm{\Phi_{RCS}}$ be a CRD matrix that uses $\mathrm{\varepsilon_{RCS}}$ as the modulating sequence: $\mathrm{\Phi_{RCS}} = \mathrm{HDF}$ where $\mathrm{D}$ contains $\mathrm{\varepsilon_{RCS}}$ on its diagonal.  It is convenient to rewrite the entries of $\Delta$, given in \eqref{eq:delta_def}, in this case as
$$\Delta_{\alpha\omega} = \sum_{j,k\neq 0}\eta_{j(j+k)}f^*_{j\alpha}f_{(j+k)\omega}\mathbb{E}[\varepsilon_j\varepsilon_{j+k}].$$
To calculate $|||\Delta|||$, it will be convenient to focus on the Gram matrix $\mathrm{\Lambda} = \Delta^*\Delta$, which has entries
\begin{equation}\label{eq:delta-gram-matrix}
  \mathrm{\Lambda}_{\alpha,\omega} = \begin{cases} \frac{W}{\ell}\sum_{j=0}^{\ell-1}e^{-\pi\imath qj}\hat{F}(j,\omega)\hat{F}^*(j,\alpha), & \omega-\alpha = \frac{W}{\ell}q \\
  0, & \text{otherwise} \end{cases}
\end{equation}
where
\begin{equation}\label{eq:windowed-spectrum}
  \hat{F}(j,\omega) = \sum_{m\neq0}\eta_{j(j+m)}f_{m\omega}\mathbb{E}[\varepsilon_{j}\varepsilon_{j+m}]
\end{equation}
and $q = 0,\pm 1,...,\pm (\ell-1)$.

We bound $|||\mathrm{\Delta}|||$ by studying the entries of $\mathrm{\Lambda}$.  To do this, recall from the definition of the spectral norm that for a matrix $\mathrm{A}$ we have $\lVert\mathrm{A}|_{\Omega'\times\Omega'}\rVert \leq \lVert\mathrm{A}|_{\Omega\times\Omega}\rVert$ for any $\Omega' \subset \Omega$.  We can therefore lower bound $|||\Delta|||$ in this case by using $\Omega$ such that $|\Omega| = 1$, i.e.,  $S=1$.  For $S=1$, $|||\Delta|||$ is the square root of the maximum entry on the diagonal of $\mathrm{\Lambda}$.  Applying \eqref{eq:rcs-definition} to the autocorrelation in \eqref{eq:windowed-spectrum}, it is straightforward to show that
\begin{equation}\label{eq:gram-matrix-diag-entries}
 \mathrm{\Lambda}_{\omega,\omega} = \frac{W}{d+1}\sum_{j=0}^{d}\sum_{\substack{m=-j \\ m\neq0}}^{d-j}\sum_{\substack{k=-j \\ k\neq0}}^{d-j}f_{m\omega}^*f_{k\omega},
\end{equation}
and that \eqref{eq:gram-matrix-diag-entries} is maximized by $\omega=0$.  This results in $\Lambda_{0,0} = d^2$ and, in the case of an RCS, for any $S$ that $|||\Delta||| \geq 1$. Finally, in the context of Theorem \ref{thm:rand-sig-coherence}, recall \eqref{eq:matrixX} for the case of $\alpha = \omega$. In this case, it is easy to see that $|x_{\alpha\omega}| \geq d+1$. Theorem \ref{thm:rand-sig-coherence}, on the other hand, relies on bounding $||\mathrm{X}||_{\max}$ close to $0$ (the details are in Appendix \ref{app:random-signal-model}) and this obviously cannot be done for an RCS.

\begin{figure}[!t]
  \centering
  \subfigure[The singular values very near to zero represent poor conditioning of the submatrices of $\mathrm{\Phi_{RCS}}$.]{
    \includegraphics[width=2.6in]{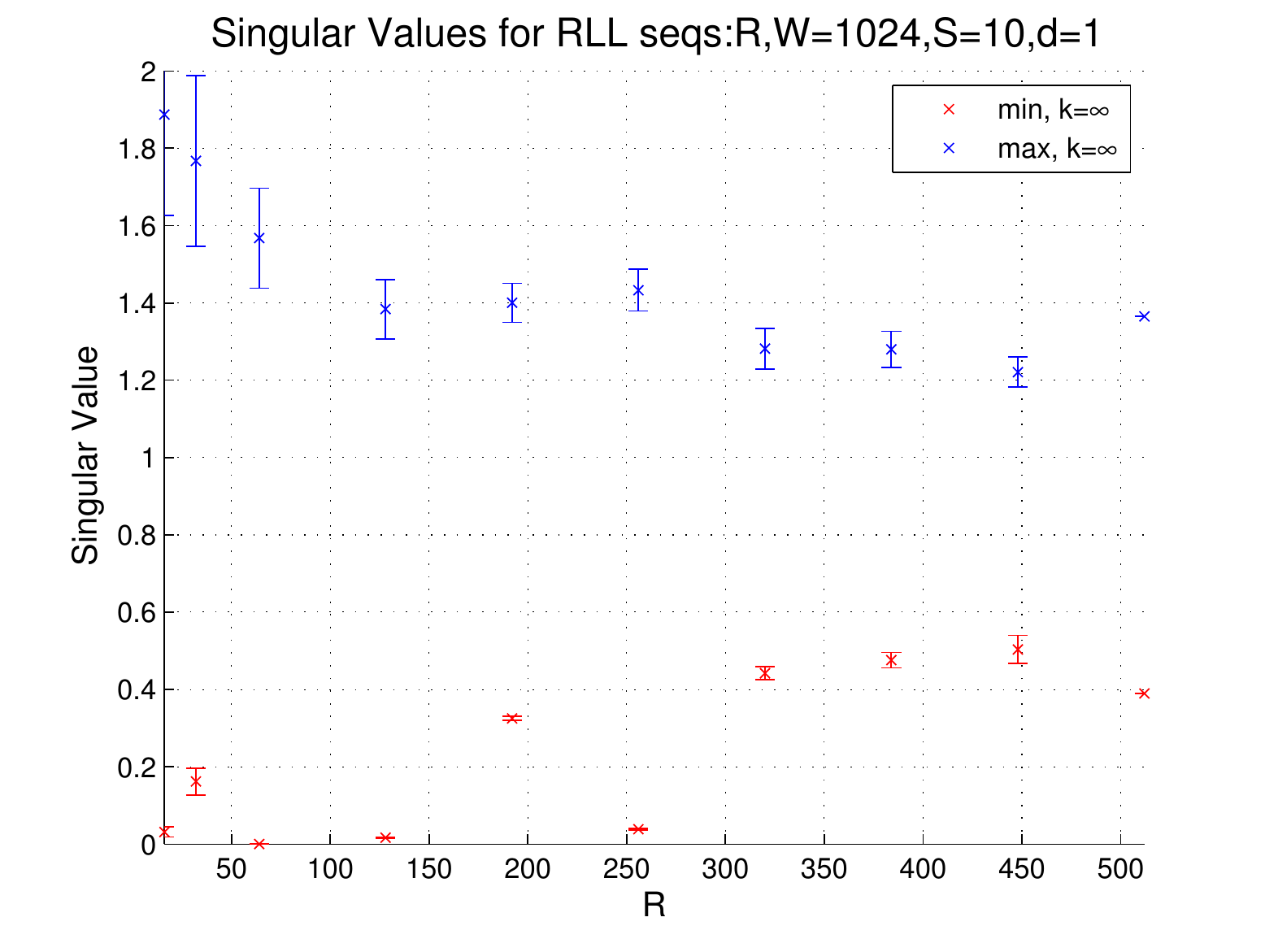}
    \label{fg:repcode_sing_val}
  }
  \subfigure[The singular values are bounded away from $0$ and $2$ indicating good conditioning of the submatrices of $\mathrm{\Phi_{MRS}}$.]{
    \includegraphics[width=2.6in]{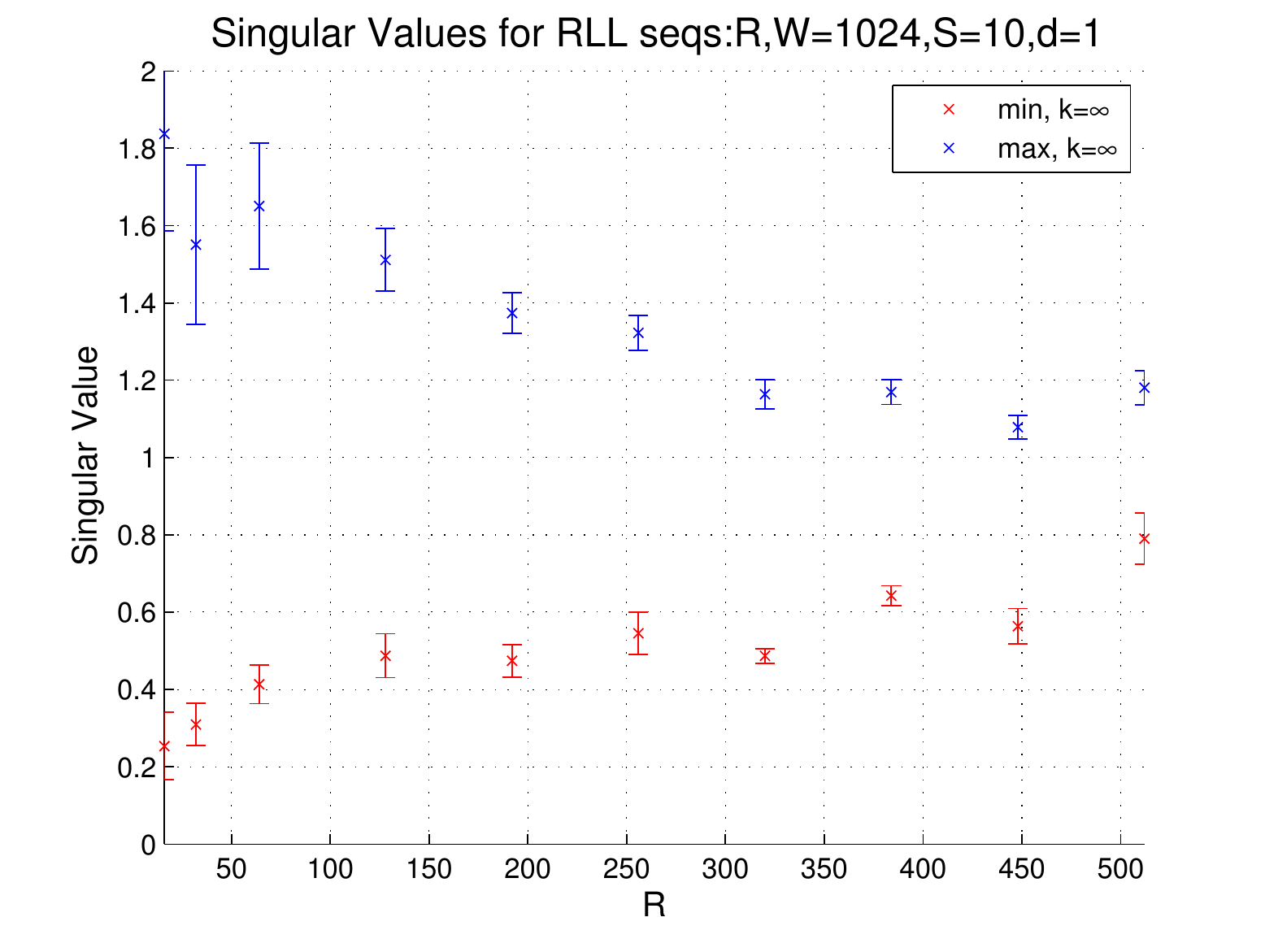}
    \label{fg:rll_sing_val}
  }
  \caption{The minimum and maximum singular values of submatrices with $10$ randomly chosen columns averaged over 1000 realizations of the measurement matrix.  The error bars represent $2$ standard deviations above and below the average value.}
  \label{fg:singular_values}

\end{figure}

We see that Theorems \ref{thm:ripcord} and \ref{thm:rand-sig-coherence} do not hold for $\mathrm{\Phi_{RCS}}$.  Although we do not have converses, we demonstrate the tightness of our theory for an RCS through numerical experiments.  For this, we calculate the minimum and maximum singular values of the submatrices over an ensemble of matrices $\mathrm{\Phi_{RCS}}$ generated using an RCS with $d=1$.  The submatrices are chosen by picking $S=10$ columns at random from $\mathrm{\Phi_{RCS}}$.  The results are presented in Fig. \ref{fg:repcode_sing_val}, where we see the minimum singular values are often at or very near zero for some values of $R$, indicating the RIP is either not satisfied or barely satisfied with an extremely small isometry constant.  Further, we show through numerical experiments in Section \ref{sec:numerical-results} that reconstruction performance is in general poor for $\mathrm{\Phi_{RCS}}$.

\section{Wide-Sense Stationary RLL Sequences}\label{sec:general-sequences}
We have seen in the previous section that $\mathrm{\Phi_{RCS}}$ does not satisfy the requirements for Theorems 1 or 2; Fig. \ref{fg:repcode_sing_val} offers further evidence that $\mathrm{\Phi_{RCS}}$ does not satisfy the RIP.  We therefore do not expect it to perform well in the sampling and reconstruction of sparse signals.  In this section, we show that a different class of RLL sequences\cite{immink98}, although more complicated than an RCS, produce measurement matrices with better conditioned submatrices and perform much better in the sampling and reconstruction of frequency-sparse signals.

We begin by examining the RIP for a modulating sequence, $\varepsilon$, that is wide-sense stationary with autocorrelation function $R_{\varepsilon}(m) = \mathbb{E}[\varepsilon_j\varepsilon_{j+m}]$.  We assume the maximum dependence distance is $\ell$, so $R_{\varepsilon}(m) = 0$ for $|m| \geq \ell$.  Under these assumptions, we want to upper bound $|||\Delta|||$.  It will be easiest to focus on the Gram matrix \eqref{eq:delta-gram-matrix}.  In this case, we can also rewrite \eqref{eq:windowed-spectrum} in terms of $R_{\varepsilon}(m)$: $\hat{F}(j,\omega) = \sum_{m\neq0}\eta_{j(j+m)}f_{m\omega}R_{\varepsilon}(m)$ which we refer to as the \emph{``windowed" spectrum} because $\eta_{j(j+m)}$ can be viewed as a ``windowing" operation on $R_{\varepsilon}(m)$.  From \eqref{eq:windowing-function}, we see that the width of the window is $W/R$, which will be quite large as $W$ increases (and $R$ scales as in (\ref{eq:thoerem1_R})).  $\hat{F}(j,\omega)$ also looks very much like the \emph{power spectrum} of $\varepsilon$: $F_{\varepsilon}(\omega) = \sum_{m}R_{\varepsilon}(m)e^{-\frac{2\pi\imath}{W}m\omega}$.  Note that $F_{\varepsilon}(\omega)$ is real-valued.  The significant differences in $\hat{F}(j,\omega)$ are the exclusion of $m=0$ in the sum, a scaling by $W^{-\frac{1}{2}}$ from $f_{m\omega}$, and the windowing by $\eta_{j(j+m)}$.  If $W/R \gg \ell$ then the windowing has negligible effect in $\hat{F}(j,\omega)$ because $R_{\varepsilon}(m) = 0$ for $|m| \geq \ell$; $\hat{F}(j,\omega)$ and \eqref{eq:delta-gram-matrix} both simplify greatly in this case.  To see this, first notice that because $\varepsilon$ is a bipolar sequence $R_{\varepsilon}(0) = 1$, and ${F}_{\varepsilon}(\omega) = \sum_{m\neq0}R_{\varepsilon}(m)e^{-\frac{2\pi\imath}{W}m\omega} + 1$ where
\begin{equation}\label{eq:reduced-spectrum}
  \tilde{F}_{\varepsilon}(\omega) \equiv \sum_{m\neq0}R_{\varepsilon}(m)e^{-\frac{2\pi}{W}m\omega} = F_{\varepsilon}(\omega) - 1.
\end{equation}
We call $\tilde{F}_{\varepsilon}(\omega)$ the \emph{reduced spectrum} of $\varepsilon$.  Under the assumption that $W/R \gg \ell$, $\hat{F}(j,\omega)$ reduces to $W^{-1/2}\tilde{F}_{\varepsilon}(\omega)$ for all $j$ except $j$ satisfying $|rW/R+j| \leq \ell$ for $r=0,\cdots,R-1$ (all but a fraction $2\ell\frac{R}{W}$).  This fraction becomes increasingly small as $W$ grows.  In this case, the entries of $\Lambda$ are approximately
\begin{align*}
  \mathrm{\Lambda}_{\alpha,\omega} &\approx \frac{1}{W}\sum_{j=0}^{W-1}e^{-\frac{2\pi\imath}{W}(\omega-\alpha)j}\tilde{F}_{\varepsilon}(\alpha)\cdotp\tilde{F}_{\varepsilon}(\omega) \notag \\
  &= \delta_{\alpha\omega}\tilde{F}_{\varepsilon}(\alpha)\cdotp\tilde{F}_{\varepsilon}(\omega) \label{eq:delta_gram_matrix}
\end{align*}
where $\delta_{\alpha\omega}$ is the Kronecker delta.  In words, $\Lambda$ is approximately a diagonal matrix with the square of the reduced spectrum on the diagonal: $\Lambda \approx \text{diag}[(\tilde{F}_{\varepsilon}(\omega))^2]$, and the eigenvalues of $\mathrm{\Lambda}$ are approximately $(\tilde{F}_{\varepsilon}(\omega))^2$.  Consequently, the singular values of $\Delta$ are approximately $|\tilde{F}_{\varepsilon}(\omega)|$.  We therefore have $||\Delta|| \approx \max_{\omega}|\tilde{F}_{\varepsilon}(\omega)|$.  Now, the spectral norm of a submatrix is upper bounded by the spectral norm of the matrix, so we finally obtain
\begin{equation}\label{eq:triplebarnorm_bound}
  |||\Delta||| \leq ||\Delta|| \approx \max_{\omega}|\tilde{F}_{\varepsilon}(\omega)|.
\end{equation}
We now have a way to estimate whether or not a stationary $\varepsilon$ is well-suited for use within the CRD.  A stationary $\varepsilon$ whose spectrum is bounded within $(0,2)$ is good;  one with $F_{\varepsilon}(\omega) = 1$ $\forall \omega$ is best.

We now present some examples to make this discussion clearer.  First, consider an independent (unconstrained) $\varepsilon$, such as the one used in the RD.  In this case, $F_{\varepsilon}(\omega) = 1$ and $\tilde{F}_{\varepsilon}(\omega) = 0$ $\forall \omega$.  The Gram matrix exactly disappears ($\Lambda = 0$) and $\Delta = 0$ confirming our previous discussion.  Next, we consider the RLL sequences described in \cite{tang70} and \cite{immink98}.  To understand how well these sequences will work in the CRD, we need to calculate the power spectrum of sequences generated from the Markov chain in Fig. \ref{fg:markovchain}.

\subsection{Power Spectrum of Markov Chain RLL Sequences}
To begin, we explicitly describe the RLL sequences in \cite{immink98}.

\begin{figure}[!t]
  \centering

  \setlength{\unitlength}{0.22in}
  \begin{picture}(13,6)
    \put(0,5.2){\circle{1.5}}
    \put(-0.2,5.0){\makebox{$1$}}
    \put(0.8,5.2){\vector(1,0){1.0}}
    \put(1.0,5.7){\makebox{\tiny$+1$}}
    \put(2.7,5.2){\circle{1.5}}
    \put(2.5,5.0){\makebox{$2$}}
    \put(3.7,5.2){\vector(1,0){1.0}}
    \put(3.9,5.7){\makebox{\tiny$+1$}}
    \put(5,5.6){\makebox{$\cdots$}}
    \put(6,5.2){\vector(1,0){1.0}}
    \put(6.2,5.7){\makebox{\tiny$+1$}}
    \put(8,5.2){\circle{1.5}}
    \put(7.8,5.0){\makebox{$d$}}
    \put(8.7,5.2){\vector(1,0){1.0}}
    \put(8.9,5.7){\makebox{\tiny$+1$}}
    \put(10,5.0){\makebox{$\cdots$}}
    \put(11,5.2){\vector(1,0){1.0}}
    \put(11.2,5.7){\makebox{\tiny$+1$}}
    \put(13,5.2){\circle{1.5}}
    \put(12.8,5.0){\makebox{$k$}}

    \put(8.0  ,4.4){\vector(4,-3){4.6}}
    \put(9.3  ,3.7){\makebox{\tiny$+1$}}
    \put(11.0,3.0){\makebox{$\cdots$}}
    \put(13.0,4.4){\vector(0,-1){3.4}}
    \put(13.3,3.7){\makebox{\tiny$+1$}}

    \put(0         ,0.2){\circle{1.5}}
    \put(-0.3    ,0.0){\makebox{$2k$}}
    \put(1.8     ,0.2){\vector(-1,0){1.0}}
    \put(1.0     ,0.7){\makebox{\tiny$-1$}}
    \put(2.2     ,0.0){\makebox{$\cdots$}}
    \put(5.1     ,0.2){\circle{1.5}}

    \put(4.5     ,0.0){\makebox{\scriptsize$k+d$}}
    \put(9.6    ,0.0){\makebox{\scriptsize$k+2$}}
    \put(12.4,0.0){\makebox{\scriptsize$k+1$}}

    \put(4.3     ,0.2){\vector(-1,0){1.0}}
    \put(3.5     ,0.7){\makebox{\tiny$-1$}}
    \put(7        ,0.2){\vector(-1,0){1.0}}
    \put(6.2    ,0.7){\makebox{\tiny$-1$}}
    \put(7.2    ,0.0){\makebox{$\cdots$}}
    \put(10.2  ,0.2){\circle{1.5}}
    \put(9.3     ,0.2){\vector(-1,0){1.0}}
    \put(8.7    ,0.7){\makebox{\tiny$-1$}}
    \put(12.1 ,0.2){\vector(-1,0){1.0}}
    \put(11.4 ,0.7){\makebox{\tiny$-1$}}
    \put(13    ,0.2){\circle{1.5}}

    \put(5.1,1.0){\vector(-4,3){4.6}}
    \put(4.8,1.5){\makebox{\tiny$-1$}}
    \put(1.5,2.0){\makebox{$\cdots$}}
    \put(0.0,1.0){\vector(0,1){3.4}}
    \put(0.3,1.5){\makebox{\tiny$-1$}}
  \end{picture}

  \caption{State diagram of the Markov chain generating an MRS (see Definition \ref{df:GRS}).  The transition probabilities are symmetric in the sense that $p_{(i+k)(j+k)} = p_{ij}$ where the sum is taken modulo $2k$.  The top half outputs the symbol +1 while the bottom half outputs -1.}
  \label{fg:markovchain}

\end{figure}
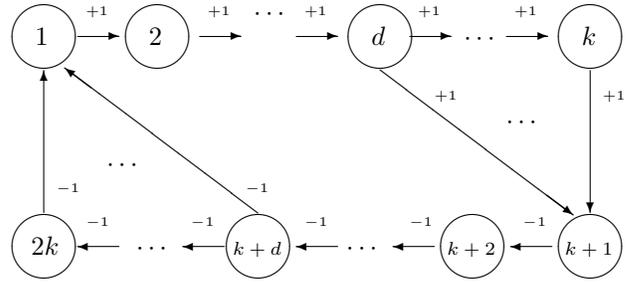

\begin{figure}[!t]
  \centering

  \includegraphics[width=2.6in]{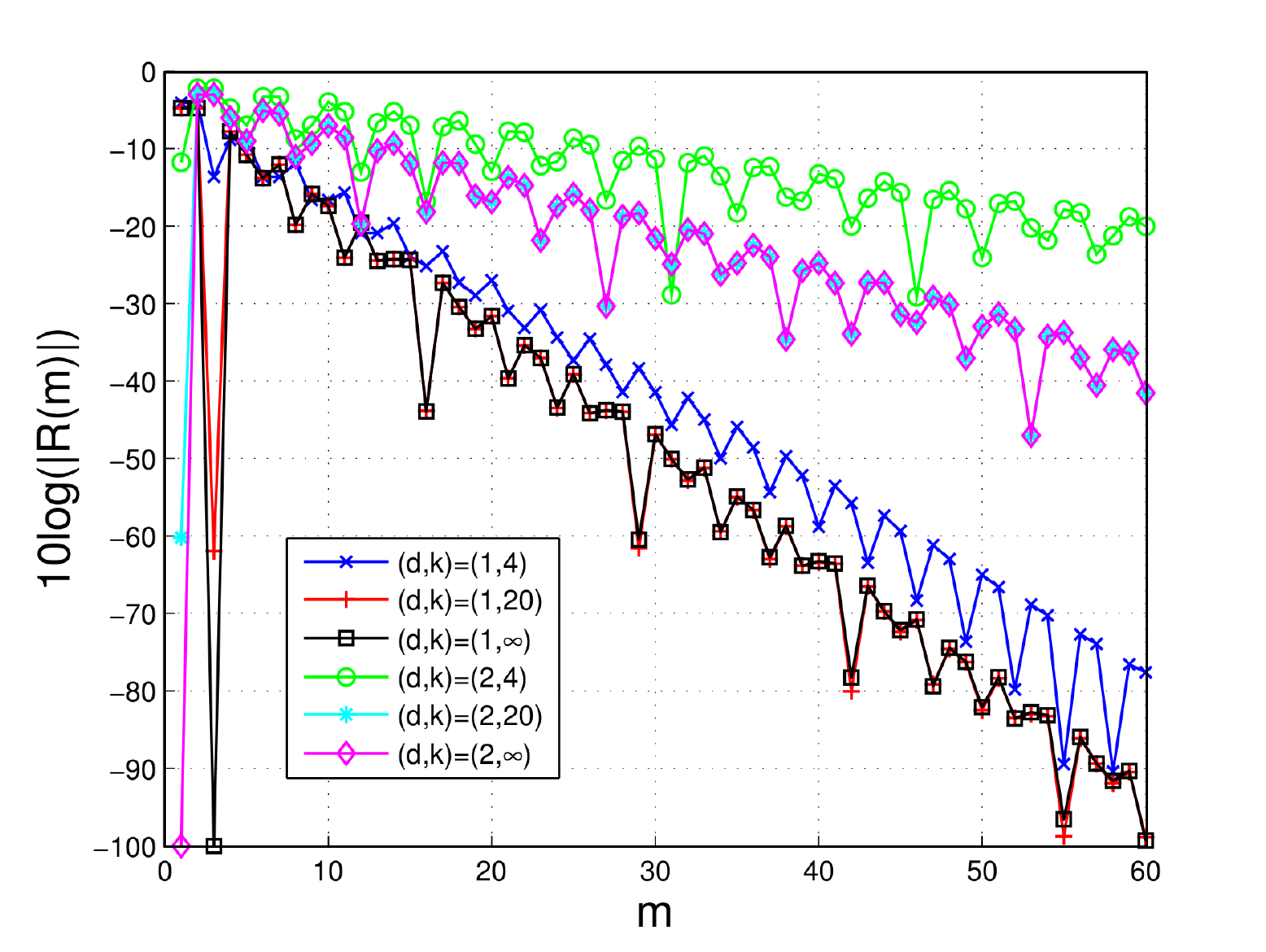}
  \caption{Log-magnitude plot of the autocorrelation of an MRS. The autocorrelation experiences geometric decay as $m\to\infty$.  The rate of decay is primarily dependent on $d$.}
  \label{fg:rll_autocorr}

\end{figure}

\begin{definition} \label{df:GRS}
We call a $(d,k)$-constrained RLL sequence that is generated from the Markov chain whose state diagram is found in Fig. \ref{fg:markovchain} a \emph{Markov-generated RLL Sequence} (MRS).  Denote such a sequence as $\mathrm{\varepsilon_{MRS}} = [\varepsilon_0, \cdots, \varepsilon_{W-1}]$ with $\varepsilon_k \in \{+1,-1\}$.  The transition probabilities are defined by the matrix $\mathrm{P} = [p_{ij}]$ where $p_{ij}$ is the probability of transitioning from state $i$ to state $j$.  The $p_{ij}$ also satisfy $p_{(i+k)(j+k)} = p_{ij}$ where the sum is modulo $2k$.  $\mathrm{P}$ is of course a stochastic matrix with rows summing to 1.  The average of the symbols output from each state $i$ are collected in the vector $\mathrm{b} = \{b_i\}$.  The stationary distribution of the states is denoted by $\pi = [\pi_i]$ and satisfies $\pi^T = \pi^T\mathrm{P}$.
\end{definition}

Having defined these MRS, we have from \cite{bilardi83} that their autocorrelation function is
$$R_{\varepsilon}(m) = \mathrm{a}^T\mathrm{P}^m\mathrm{b}$$
where $\mathrm{a}^T = \mathrm{b}^T\cdotp\text{diag}[\pi_1,\cdots,\pi_{2k}]$ and $R_{\varepsilon}(-m) = R_{\varepsilon}(m)$. To understand the performance of an MRS within the CRD, we need to understand the behavior of $R_{\varepsilon}(m)$ as $m$ increases.  Since $\mathrm{P}$ is a stochastic matrix, we can make use of the theory of non-negative matrices to understand how $R_{\varepsilon}(m)$ behaves.  First note that $\mathrm{b}$ is orthogonal to $\mathrm{w}$, where $\mathrm{w} = [1, 1, \cdots, 1]^T$, and that $\mathrm{a}^T\mathrm{b} = 1$.  Since $\mathrm{P}$ is a stochastic matrix, its second largest eigenvalue $\lambda_2$ satisfies $\lambda_2 < 1$.  Making use of \cite[Theorem 8.5.1]{hornandjohnson}, we can bound the autocorrelation (in magnitude) as
\begin{equation}\label{eq:autocorr_bound}
  |R_{\varepsilon}(m)| = |\mathrm{a}^T\mathrm{P}^m\mathrm{b}| \leq \lambda_2^m.
\end{equation}
We see that $|R_{\varepsilon}(m)|$ experiences geometric decay, at a rate determined by $\lambda_2$.  This is confirmed in Fig. \ref{fg:rll_autocorr} where $10\log_{10}|R_{\varepsilon}(m)|$ is plotted for several pairs $(d,k)$.  Notice that the rate of decay (in magnitude) is smaller for larger values of $d$ and larger for larger values of $k$, and the curve is roughly the same for $k=20$ and $k=\infty$.  These facts can be directly tied to the eigenvalues of $\mathrm{P}$ in each case.

To evaluate the performance of an MRS within the CRD, we must evaluate the MDD and the matrix $\Delta$.  Looking first at the MDD, we use (\ref{eq:autocorr_bound}) and the fact that $\lambda_2 < 1$ to establish that $\lim_{m\to\infty}|R_{\varepsilon}(m)| = 0$ and, hence, for any $\xi > 0$, $|R_{\varepsilon}(m)|$ $ < \xi$ for all $m \geq M$ where $M = M(\xi) < \infty$.  Though we cannot guarantee that an MRS becomes completely uncorrelated for a finite $M$, we can make $\xi$ as small as we want so that the sequence is nearly uncorrelated for large enough $M$.  In this case, we can take the MDD to be $\ell \approx M(\xi)$ for some small $\xi$.  In other words, an MRS satisfies the setting of Theorem \ref{thm:ripcord} in an approximate sense.  We believe this is justified from an engineering perspective because the correlation can be made very small; the numerical experiments in Section \ref{sec:numerical-results} add further justification to this.

Next, we estimate $|||\Delta|||$ from the reduced spectrum of the MRS.  Using \eqref{eq:triplebarnorm_bound} and $\ell \approx M(\xi)$ from above, we have that $|||\Delta||| \leq ||\Delta|| \approx \max_{\omega}|\tilde{F}_{\varepsilon}(\omega)|$ where $\tilde{F}_{\varepsilon}(\omega)$ is the reduced spectrum.  We emphasize again that $\xi$ can be made as small as we like at the expense of a larger $\ell$.  Consequently, it can be argued that an MRS that satisfies $$\max_{\omega}|\tilde{F}_{\varepsilon}(\omega)| < 1$$ leads to a matrix $\mathrm{\Phi_{MRS}}$ that approximately satisfies the RIP by virtue of Theorem \ref{thm:ripcord}.

Turning to Theorem \ref{thm:rand-sig-coherence}, we must show that $||\Gamma||$ is bounded independent of $W$.  It is easy to show that for $i < j$, $$\gamma_{ij} = \sqrt{|R_{\varepsilon}(j-i)|/2} \leq \sqrt{\lambda^{j-i}_2/2}$$ for an MRS.  It is then also straightforward to show (see, e.g., the discussion after \cite[Proposition 1]{samson00}) that
$$||\Gamma|| \leq 1/\sqrt{2}(1-\lambda_2^{1/2}).$$
Since $||\Gamma||$ is independent of $W$, we can make $W$ large enough so that \eqref{eq:W-constraint-gamma} is satisfied and Theorem \ref{thm:rand-sig-coherence} is approximately satisfied.

\begin{figure}[!t]
  \centering
  \subfigure[MRS with $d=1$ and $k=20$]{
    \includegraphics[width=2.6in]{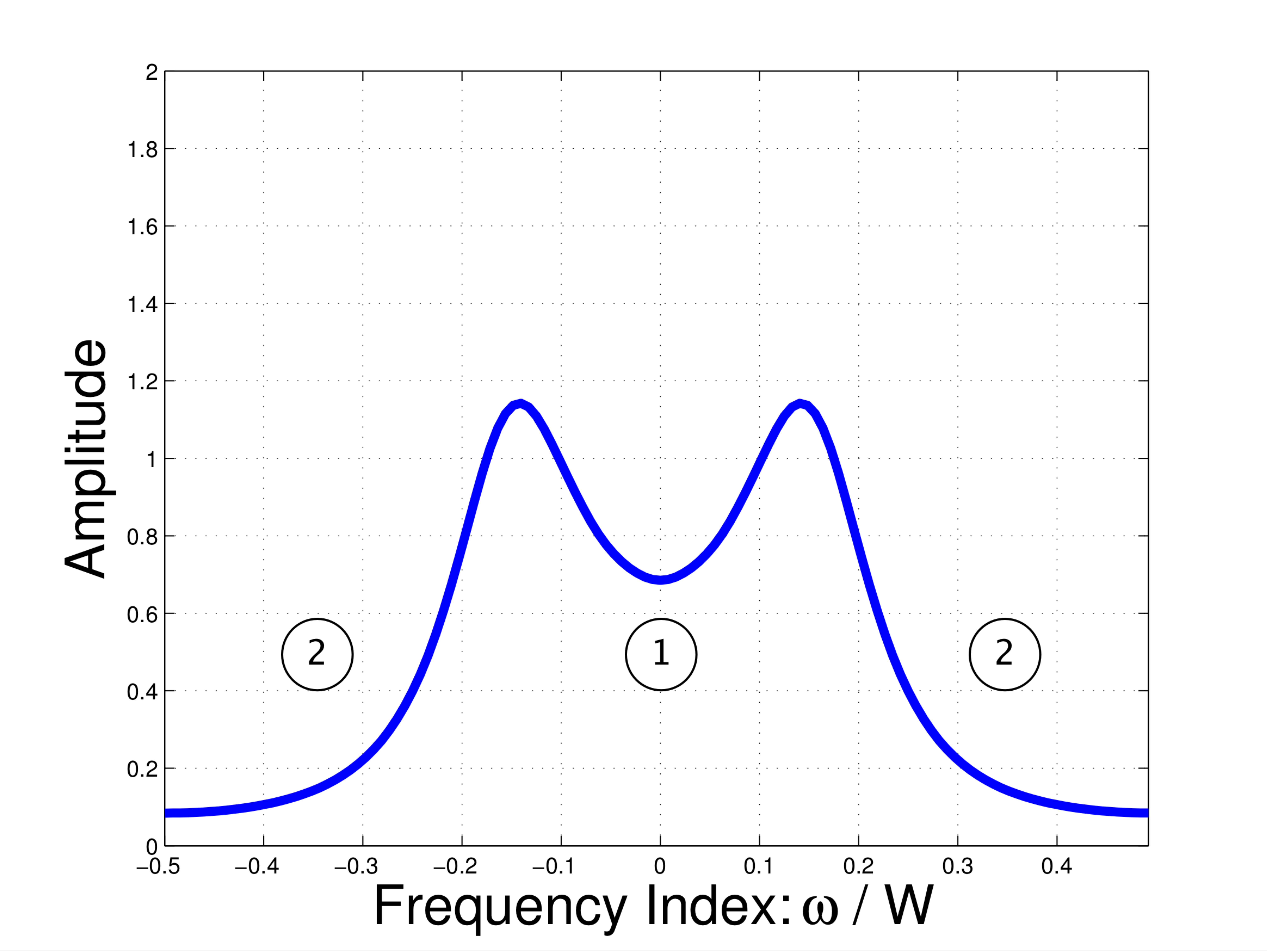}
    \label{fg:rll_spectrum}
  }
  \subfigure[Rademacher seqeunce]{
    \includegraphics[width=2.6in]{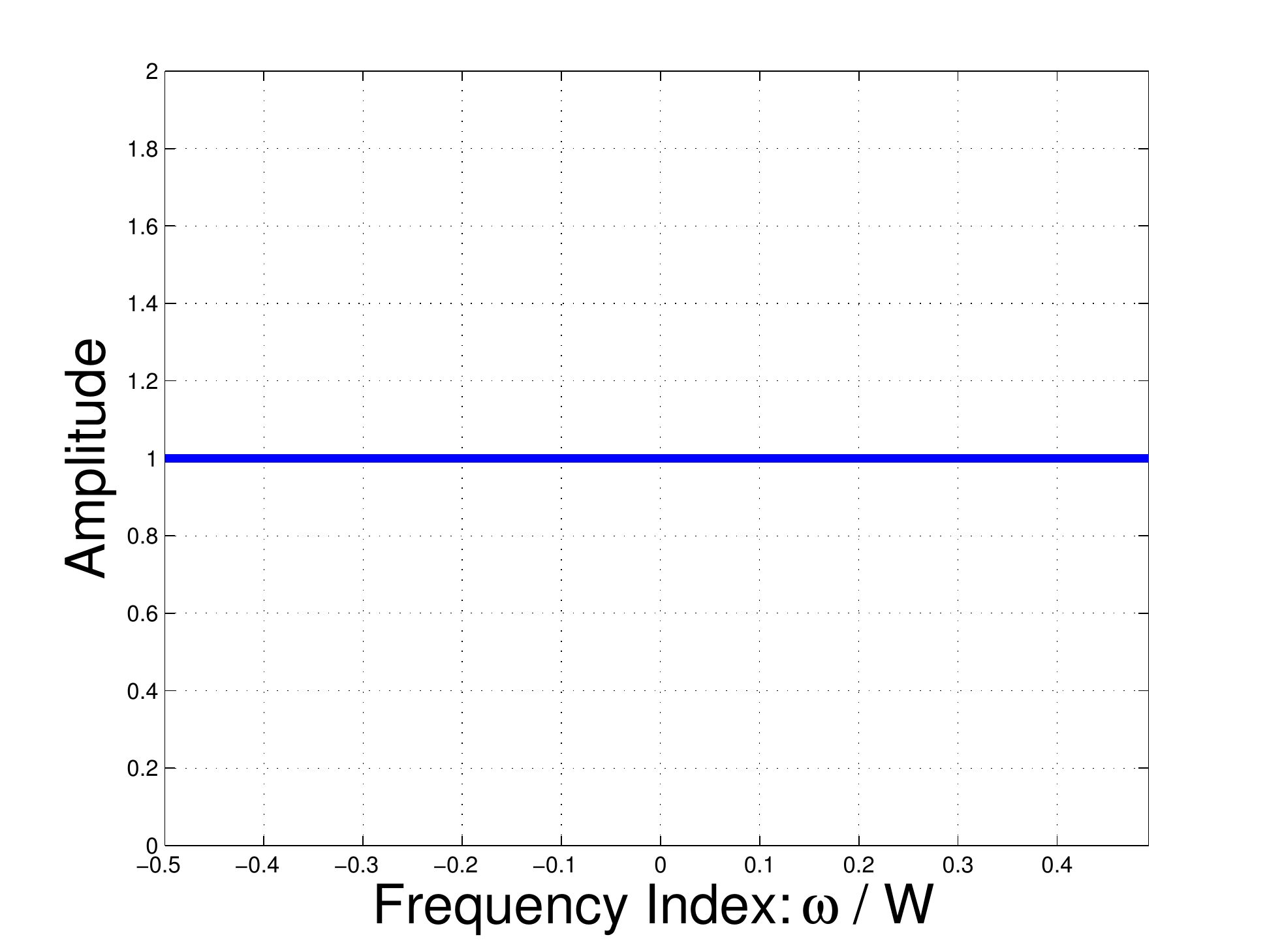}
    \label{fg:rd_spectrum}
  }
  \subfigure[RCS with $d=1$]{
    \includegraphics[width=2.6in]{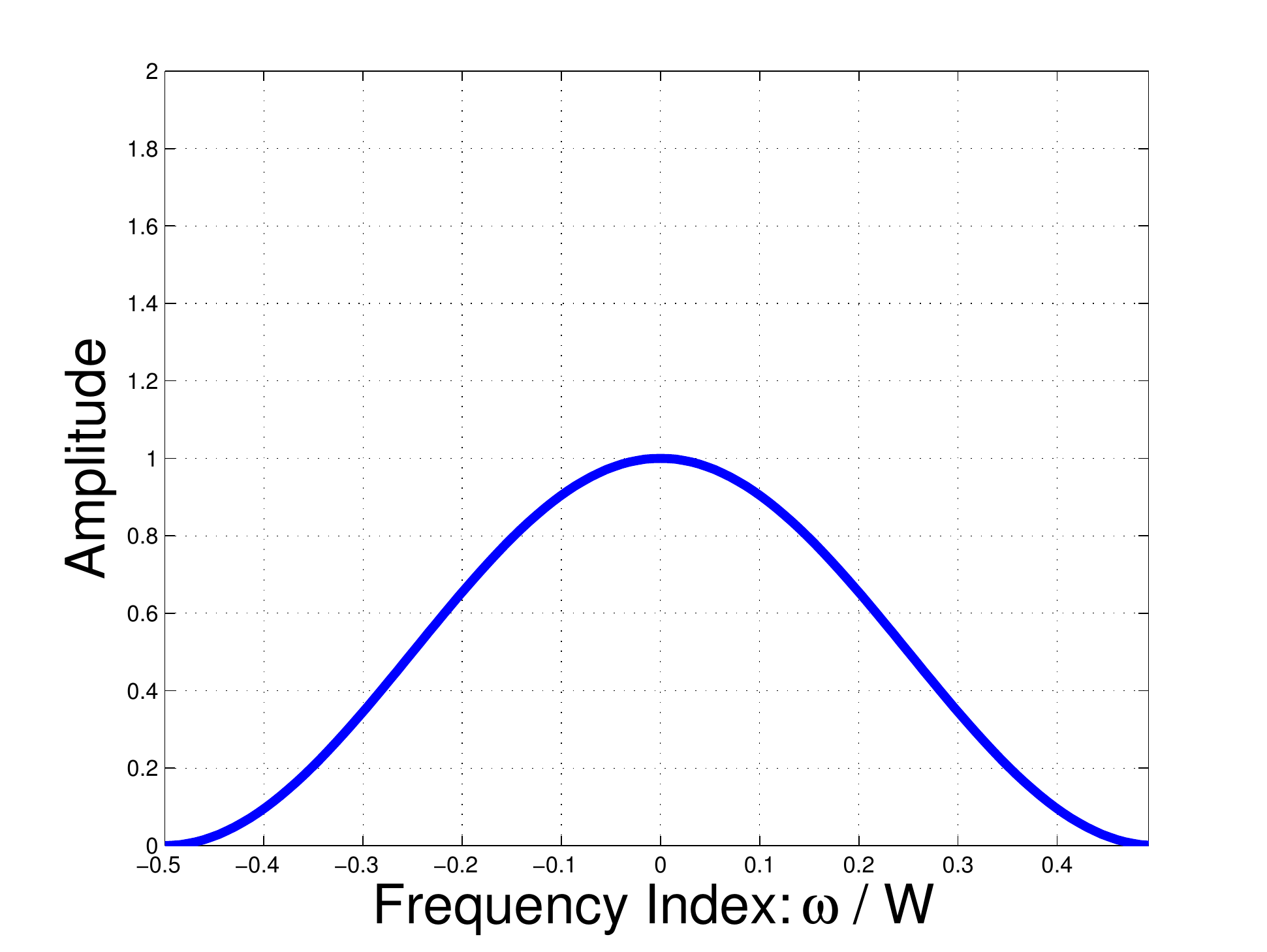}
    \label{fg:rcs_d1_spectrum}
  }
  \caption{Power spectrum of an MRS, a Rademacher sequence, and an RCS.  For the MRS, signals in region $\mathbf{1}$ get more energy in the measurements than signals in region $\mathbf{2}$.  The Rademacher sequence is ideal for sampling any frequency sparse signal.  An RCS is not well suited for sampling signals with any high frequency content.}
  \label{fg:spectrums}
\end{figure}

Our argument for the use of an MRS within the CRD makes use of some approximations.  To demonstrate the validity of these approximations, we consider an MRS with $(d,k) = (1,20)$.  The spectrum of this MRS is shown in Fig. \ref{fg:rll_spectrum}.  From this figure, we see that $\max_{\omega}|\tilde{F}_{\varepsilon}(\omega)| \approx 0.9$ corresponding to $\omega = \pm 0.5$.  Our theory, therefore, predicts that the matrix $\mathrm{\Phi_{CRD}}$ in this case satisfies the RIP.  To verify this, we calculate the average minimum and maximum singular values of the submatrices of $\mathrm{\Phi_{CRD}}$ and present the results in Fig. \ref{fg:rll_sing_val} for submatrices containing $10$ columns.  We see that as $R$ decreases, the singular values approach $0$ and $2$ but remain bounded away from them.  In Section \ref{sec:numerical-results}, we carry out numerical reconstruction experiments to further validate our theory.

\section{Random Demodulator vs. Constrained Random Demodulator: Numerical Results}\label{sec:numerical-results}
In this section we numerically contrast the performance of the RD with that of the CRD.  In the case of the CRD, we focus on measurement matrices built using the RCS and MRS.  The results here are obtained using the YALL1 software package, an $\ell_1$-solver using alternating direction algorithms\cite{yang-zhang10}.  We first examine the use of an RCS and show that a CRD using these sequences gives unsatisfactory results.

\begin{figure}[!t]
  \center
  \subfigure[Probability of successful reconstruction over 1000 {instances} of $\mathrm{\Phi_{RCS}}$ with $d=1$]{
    \includegraphics[width=2.7in]{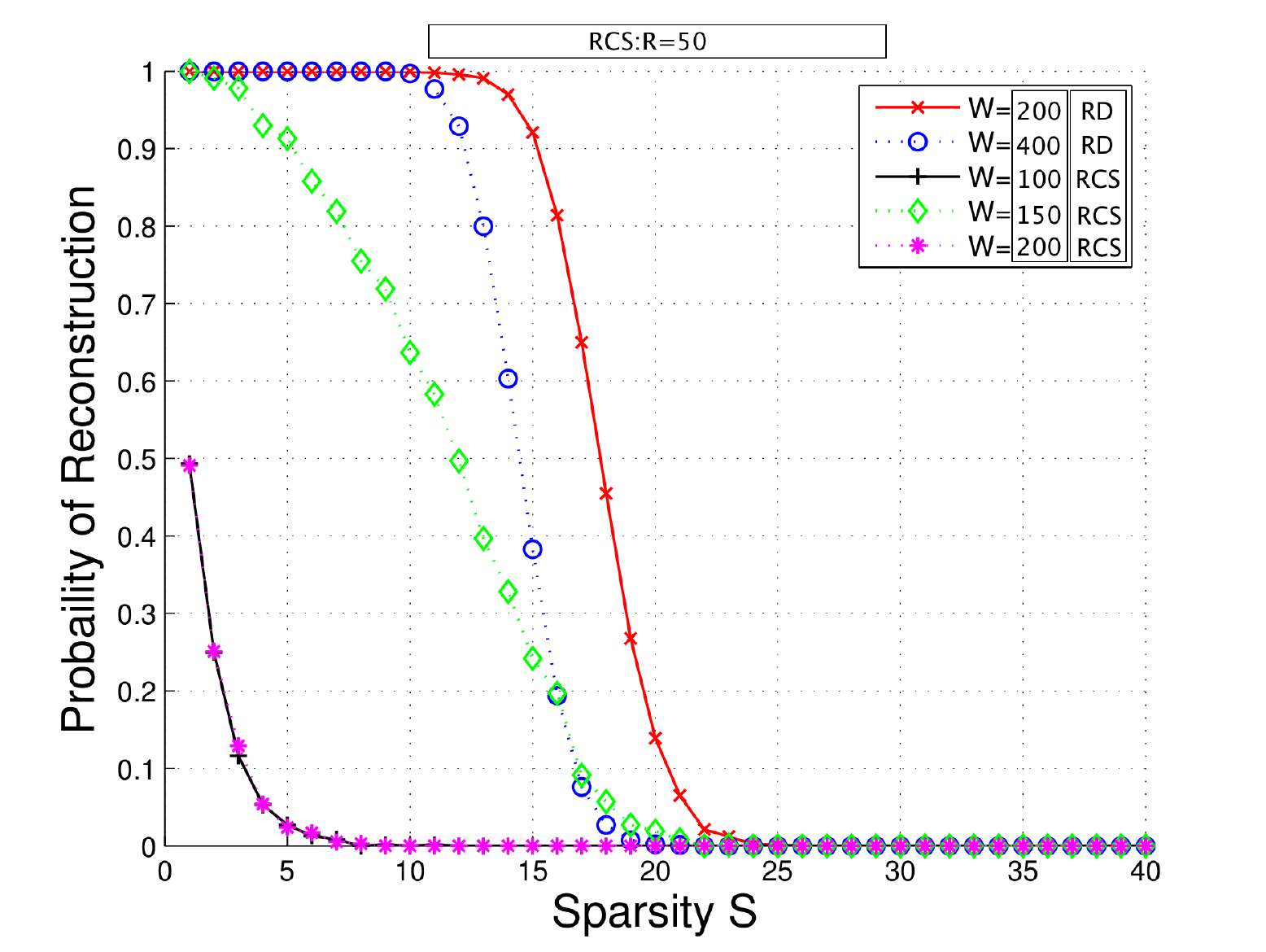}
    \label{fg:repcode_reconstruct}
  }
  \subfigure[Probability of successful reconstruction over 1000 instances of $\mathrm{\Phi_{MRS}}$ with $d = 1$ and $k = 20$]{
    \includegraphics[width=2.7in]{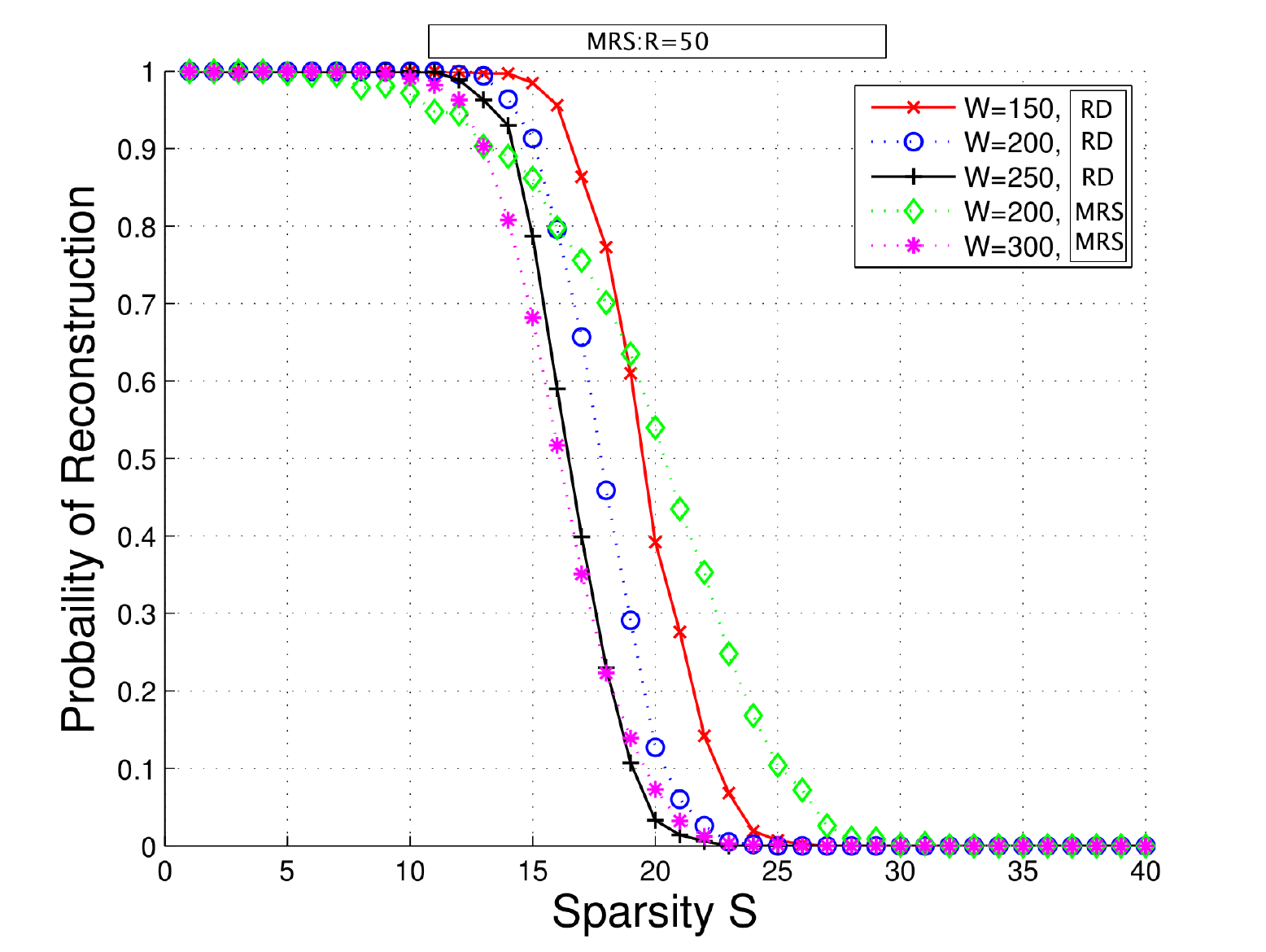}
    \label{fg:rll_reconstruct}
  }
  \caption{The RCS does not offer good performance and, in fact, fails quite often.  The MRS offers comparable performance to the Rademacher sequences of the RD.}
  \label{fg:compare_RCS_GRS}

\end{figure}

Recall that we have argued in Section \ref{sec:RCS}  that $\mathrm{\Phi_{CRD}}$ using an RCS does not satisfy the RIP.  Consequently, if we sample a sparse signal using such a measurement matrix and attempt to reconstruct it, we expect to get poor results.  This is indeed the case in our numerical experiments as shown in Fig. \ref{fg:repcode_reconstruct}.  To produce these results, we hold the sampling rate constant at $R=50$ and vary the bandwidth $W$.  It is particularly noteworthy that sampling and reconstruction fail most of the time at $W=100$ and $W=200$.  Note that the RCS performs relatively better at $W = 150$, owing to the splitting of some repeated entries of the RCS between successive rows of the composite matrix $\mathrm{HD}$.

We then examine sampling with a CRD that uses an MRS with $d=1$ and $k=20$ and show that it produces results similar to those for the RD using a Rademacher sequence.  Recall that we have argued in Section \ref{sec:general-sequences} the usefulness of RLL sequences generated from the Markov chain of Fig. \ref{fg:markovchain} in the context of the CRD.  Fig. \ref{fg:rll_reconstruct} validates this assertion and shows the empirical probability of reconstruction if we sample sparse signals with $\mathrm{\Phi_{CRD}}$ that uses these sequences.  The baseline for comparison is of course the RD.  The figure shows that the performance using an MRS is very similar to the performance using the Rademacher sequences of the RD.  In fact, the CRD allows us to tradeoff between sparsity, bandwidth, and recovery success.  In particular, if we concentrate on the RD curve at $W=250$ and the CRD curve at $W=300$, we see that at a $90\%$ success rate, we only pay a sparsity penalty of $2$ ($\approx 13\%$) by using the CRD.  At the same time, however, we have gained an advantage in bandwidth, $W$, of $20\%$.  Comparing the CRD curve at $W=300$ to the RD curve at $W=150$ we see that at a $90\%$ success rate, we incur approximately a $28\%$ sparsity penalty for a $100\%$ increase in bandwidth.  Other tradeoffs can be seen at different success rates, but it is reasonable to argue that most applications will operate A/D converters in the high success rate regions.  At lower success rates, the advantage is even greater for the CRD.  While our analysis concentrates on a high success rate, analysis at lower success rates could prove useful for future work.

\section{Knowledge Enhanced Sub-Nyquist Sampling}\label{sec:kecom-sampling}
In this section, we argue that the performance of a CRD can be enhanced by leveraging a priori knowledge about the signal.  We notice two operations in Fig. \ref{fg:rd} that are central to the functioning of the RD/CRD: the modulation by the random waveform and the subsequent low-pass filtering.  The low-pass filtering operation allows the RD/CRD to operate at the sub-Nyquist rate $R$, while modulation by the random waveform---which smears the input signal tones across the spectrum, including in the low-pass region---results in a unique signature of each tone within the low-pass region.  Theorem \ref{thm:ripcord} states the sufficient conditions for uniqueness to hold for all possible input signals, and we explored in Section \ref{sec:general-sequences} how the RIP depends on the power spectrum of the random sequence.  In addition to uniqueness of each tone's signature in the low-pass region, the performance of the RD/CRD depends on the energy smeared into the low-pass region because tones with a low-energy signature will be harder to recover.

Note that the modulation by the random waveform in time is equivalent to a convolution in the frequency domain.  Therefore, the power spectrum of the random waveform tells us how much energy from each tone \emph{on average} is smeared into the low-pass region (and thus collected in the measurements).  Inspection of \eqref{eq:triplebarnorm_bound} tells us the RIP depends on the worst-case deviation from a flat spectrum.  However, if we use an MRS within the CRD and the input signal is statistically more likely to contain low frequencies, then this additional knowledge about the signal can be leveraged to improve the reconstruction averaged over many signals and random waveform realizations.  Note that this is a different ``average case" setup than the one in Theorem \ref{thm:rand-sig-coherence}.  Here, we impose a nonuniform distribution on tones in the input signal.  We show in this setting that the CRD can perform better than the RD, provided the statistical distribution of the tones is matched to the power spectrum of the MRS, because the CRD in this case will \emph{on average} smear and capture more energy from the input tones in the low-pass region of the spectrum.  In addition to the case of possessing prior knowledge about the input signal distribution, the exposition in this section is also of interest in other scenarios. Consider, for example, a spectrum sensing application in which one assigns a higher priority of detection to some regions and a lower priority of detection to other regions. Similarly, consider the case where one possesses knowledge about colored noise or narrowband interference injected into the signal. In both these settings, the CRD can be tailored through the choice of the modulating waveform to perform better than either the RD, which treats all spectral regions the same way, or a pure passband system, which completely throws away information in some spectral regions.  We term such usage of the CRD that exploits prior knowledge a \emph{knowledge-enhanced CRD}.

Note that somewhat similar ideas have been briefly explored in \cite{setti10} and \cite{setti11}, but without the explicit examination of the uniqueness of tone signatures.  Recent work on model-based compressed sensing also attempts to leverage additional a priori information in the signal model \cite{baraniuk10}, but the focus there is exclusively on the reconstruction side instead of the sampling side.

\subsection{Phase Transitions of Reconstruction Success}

\begin{figure}[!t] 
  \centering
  \subfigure[RD with a uniform distribution on the input tones.]{
    \includegraphics[width=2.45in]{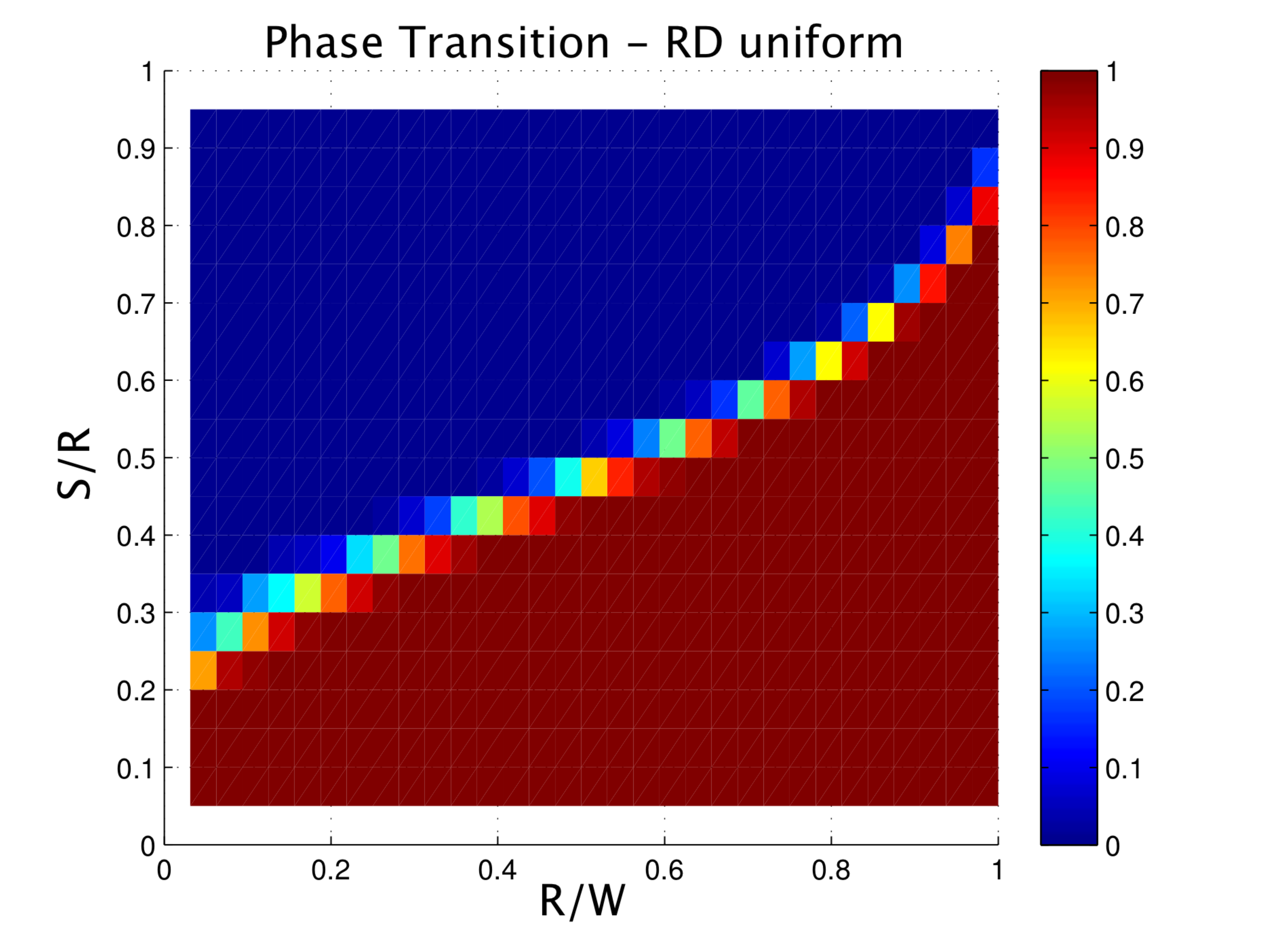}
    \label{fg:rd_phasetransition_uniform}
  }
  \subfigure[RD with a distribution on the input tones matched to the power spectrum of a $(1,20)$ RLL sequence.]{
    \includegraphics[width=2.4in]{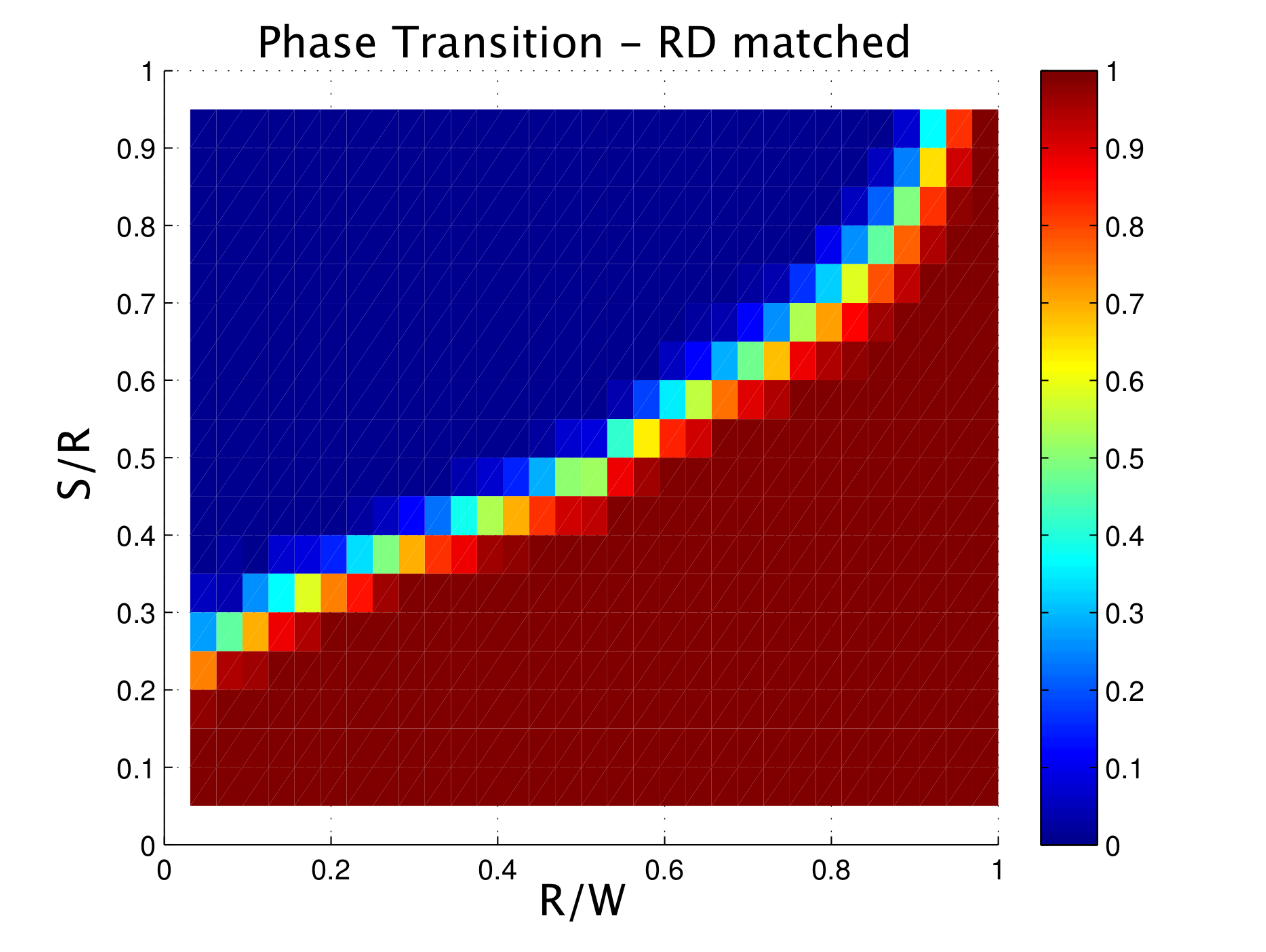}
    \label{fg:rd_phasetransition_matched}
  }
  \subfigure[CRD with a uniform distribution on the input tones.]{
    \includegraphics[width=2.45in]{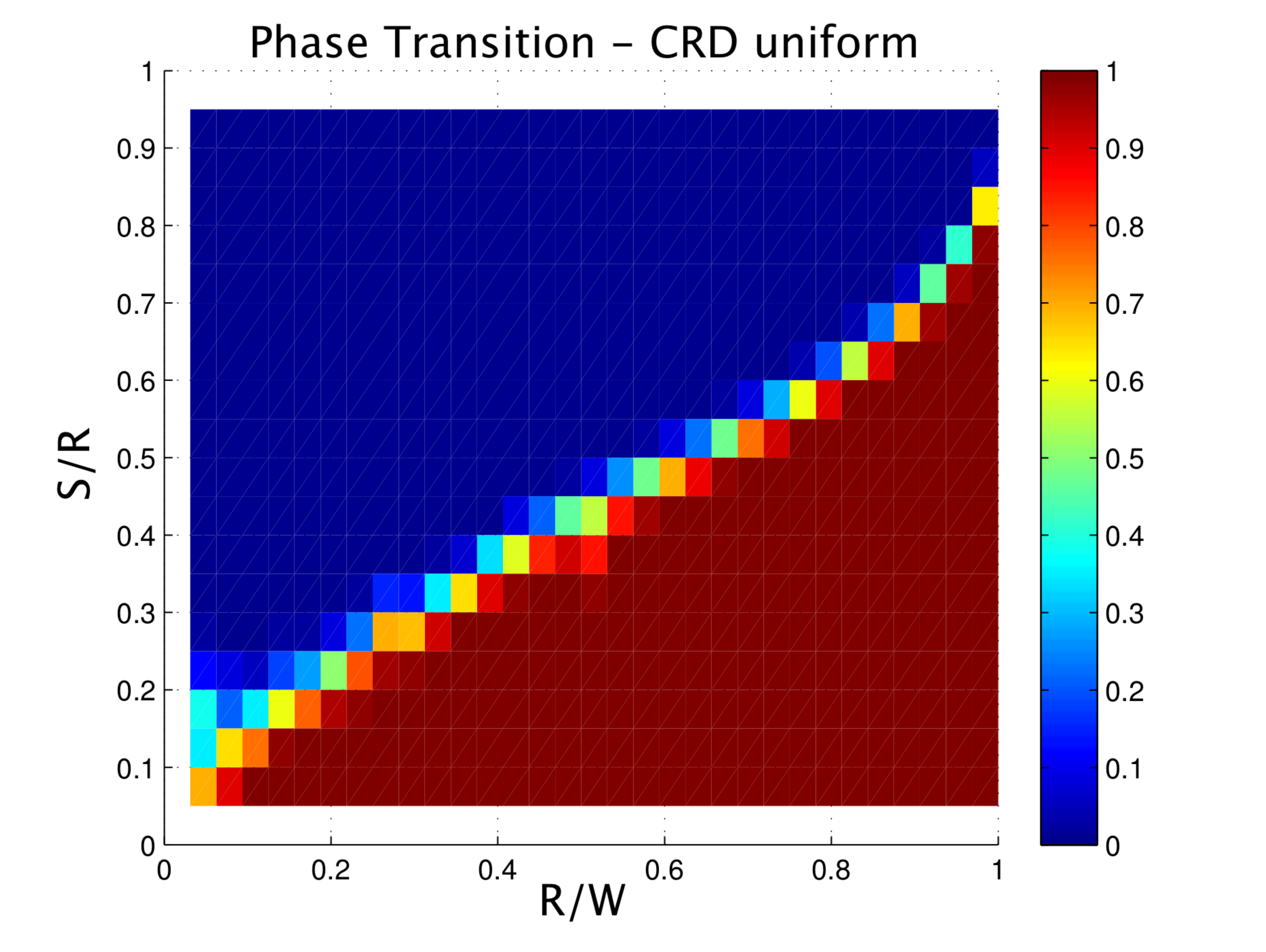}
    \label{fg:crd_phasetransition_uniform}
  }
  \subfigure[CRD with a distribution on the input tones matched to the power spectrum of a $(1,20)$ RLL sequence.]{
    \includegraphics[width=2.4in]{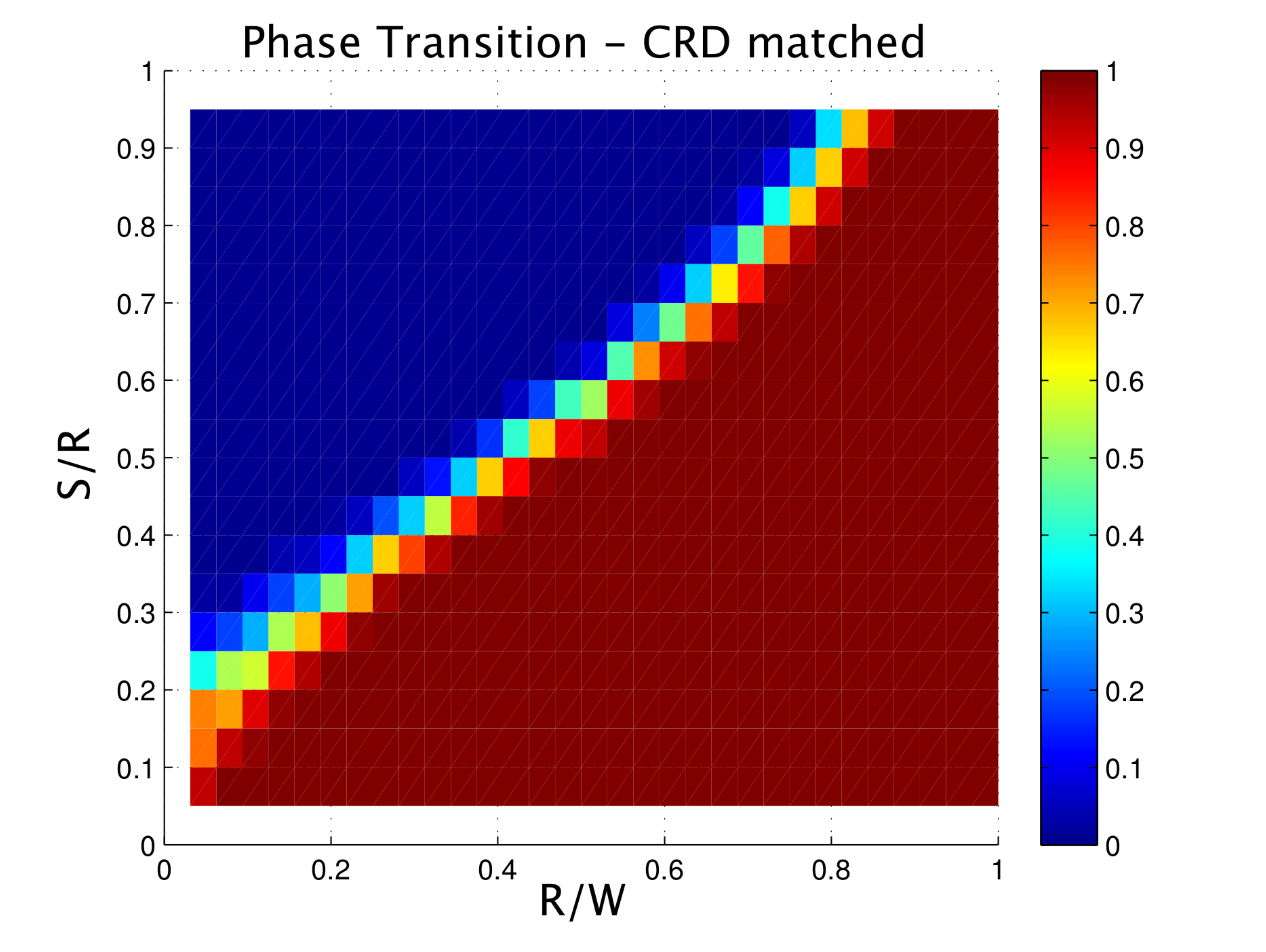}
    \label{fg:crd_phasetransition_matched}
  }
  \caption{Empirical reconstruction success as a function of $S/R$ and $R/W$.  The phase transition is the transition from 0 to 1.}
  \label{fg:phasetransition_plots}
\end{figure}

To verify our understanding of the knowledge-enhanced CRD, we have conducted extensive numerical simulations to compare reconstruction performance for signals sampled by a CRD (using an MRS) against the RD (using a Rademacher sequence).  Our focus here will be on two classes of input signals.  The first class is generated by drawing a sparse set of tones uniformly at random; the second class is generated with a distribution on the tones that matches the power spectrum of an MRS with $(d,k)=(1,20)$ (see Fig. \ref{fg:rll_spectrum}).  We also focus on two measurement matrices: the RD and the CRD using an MRS with $(d,k) =(1,20)$.  Recall, the RD uses an (unconstrained) Rademacher sequence.  The sequence is comprised of independent terms, resulting in a flat spectrum (see Fig. \ref{fg:rd_spectrum}).  Because the spectrum is flat, a Rademacher sequence will illuminate all tones equally well.  That is to say, we expect good reconstruction performance for all sparse signals.  On the other hand, the MRS used in the CRD has correlations between terms of the sequence that gives rise to the spectrum in Fig. \ref{fg:rll_spectrum}.  We see that the spectrum is close to $1$ for the low frequencies (Region $\textbf{1}$) and approximately $0.1$ at high frequencies (Region $\textbf{2}$).  If low-frequency tones are statistically more likely in the input signal, then we expect the CRD on average to capture more energy in the measurements and offer better reconstruction performance. Note, we do not consider the CRD using an RCS because we have shown in Section \ref{sec:numerical-results} that the reconstruction performance is very poor.  To understand why it is poor for an RCS, we can examine the spectra of these sequences.  An RCS is not stationary but rather cyclo-stationary, so we calculate the spectrum by averaging over the cycle period.  The resulting spectrum is shown in Fig. \ref{fg:rcs_d1_spectrum} for $d=1$.  The spectrum approaches zero at high frequencies, so we expect the CRD in this case to capture very little energy from high frequency tones in the low-pass region.  Consequently, we also expect poor reconstruction performance.

The results are displayed in Fig. \ref{fg:phasetransition_plots} for the four combinations described above: two input signal classes and two measurement matrices.  For these experiments, an RD or CRD matrix is generated using a random instance of the modulating sequence 3000 times for each point (i.e., pixel) on the plot.  The matrix is used to sample a new randomly generated $S$-sparse vector, and reconstruction of the original vector from its samples is carried out using the YALL1 software package.  Success is defined as the two vectors being equal to each other to 6 decimal places in the $\ell_{\infty}$ norm.  The results in Fig. \ref{fg:phasetransition_plots} show that the RD performs (almost) equally well for the two input signal classes.  On the other hand, the CRD performs much better for the second class of input signals.  Additionally, the CRD suffers more at very small $R/W$ ratios.

\begin{figure}[!t] 
  \centering
  \subfigure[MSE plot for a RD for signals with a uniform distribution on the tones.]{
    \includegraphics[width=2.45in]{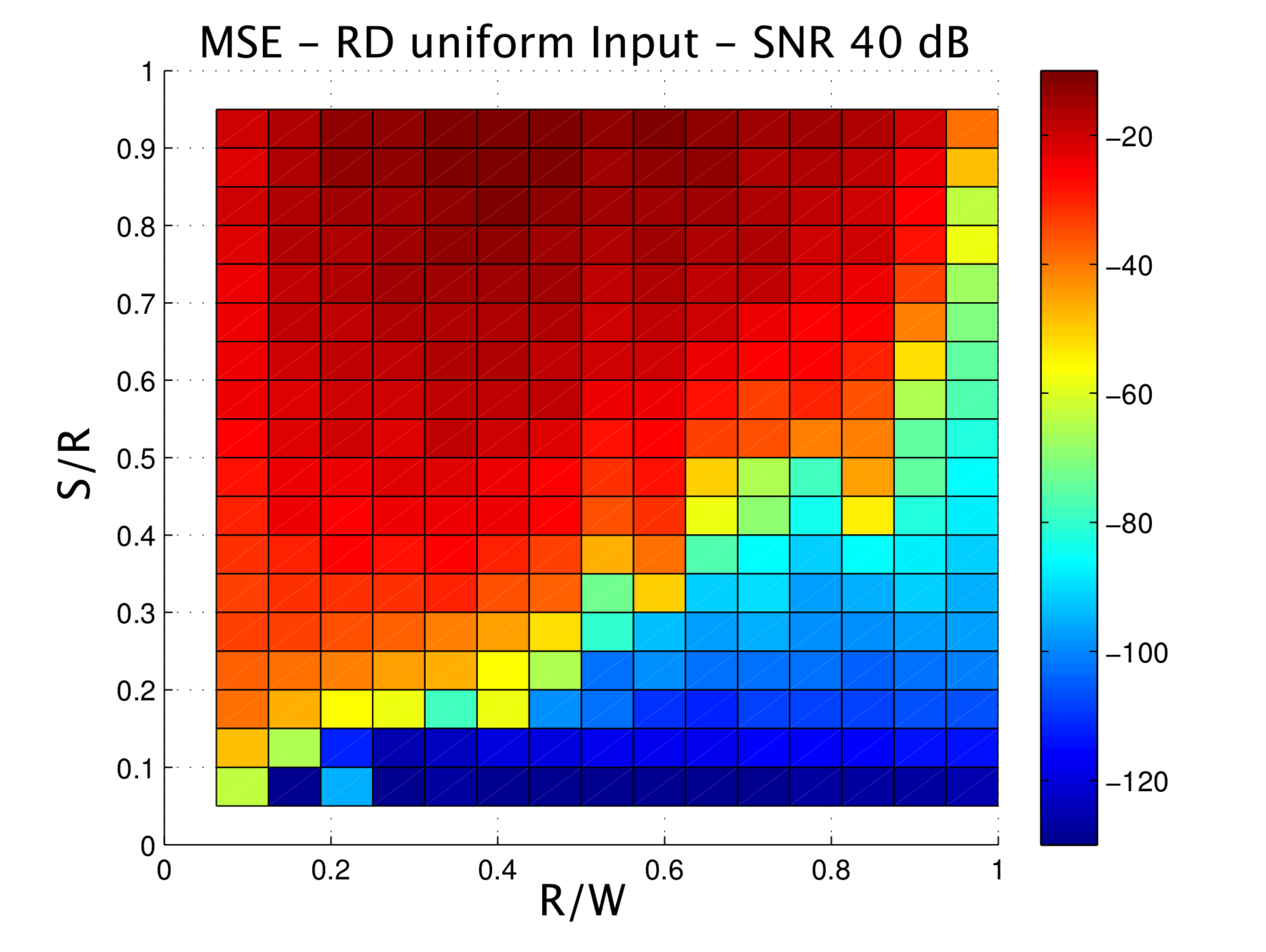}
    \label{fg:rd_MSE_uniform}
  }
  \subfigure[MSE plot for a RD for signals with a distribution on the tones that matches the spectrum in Fig. \ref{fg:rll_spectrum}.]{
    \includegraphics[width=2.45in]{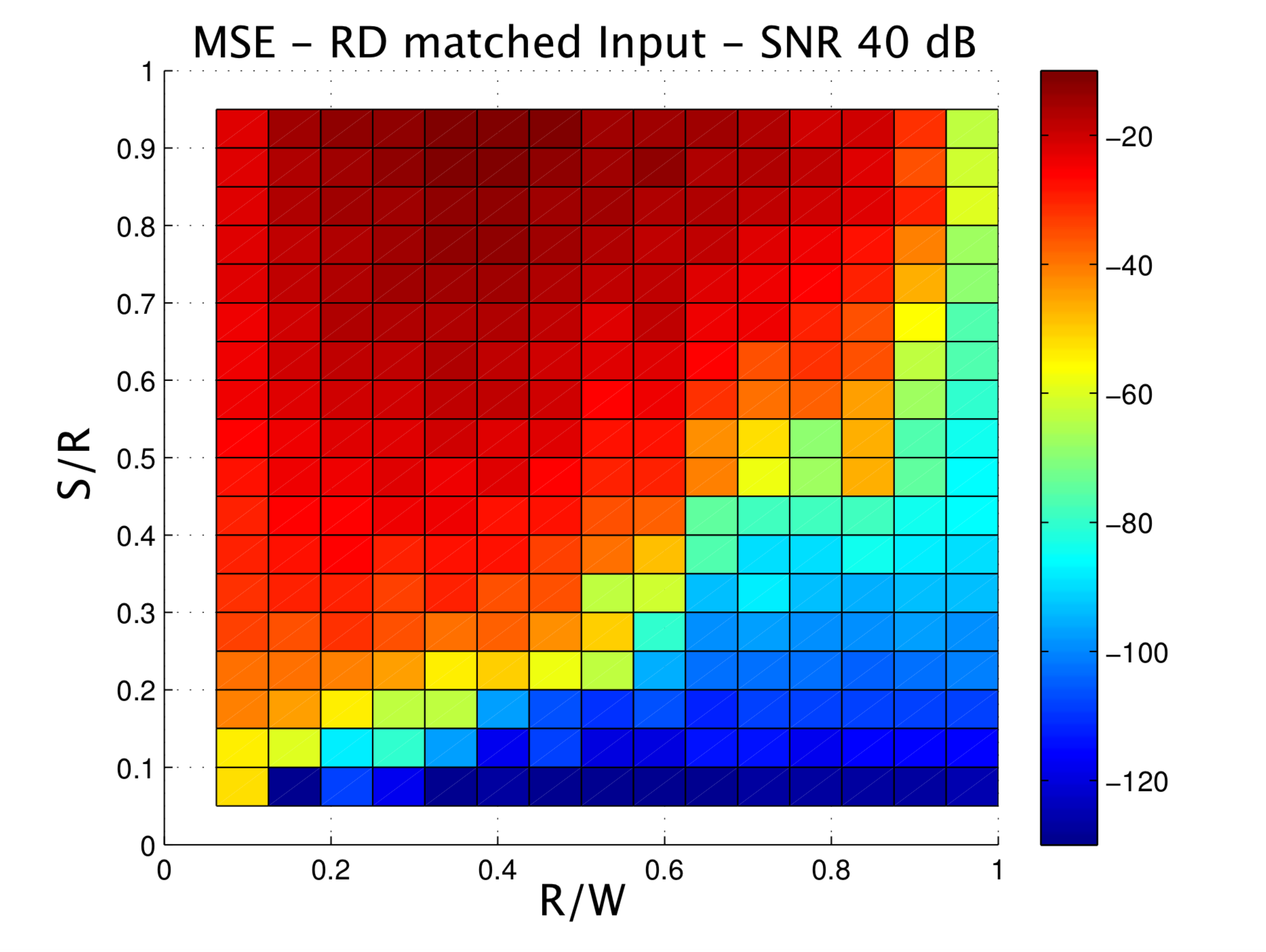}
    \label{fg:rd_MSE_matched}
  }
  \subfigure[MSE plot for a CRD for signals with a uniform distribution on the tones.]{
    \includegraphics[width=2.45in]{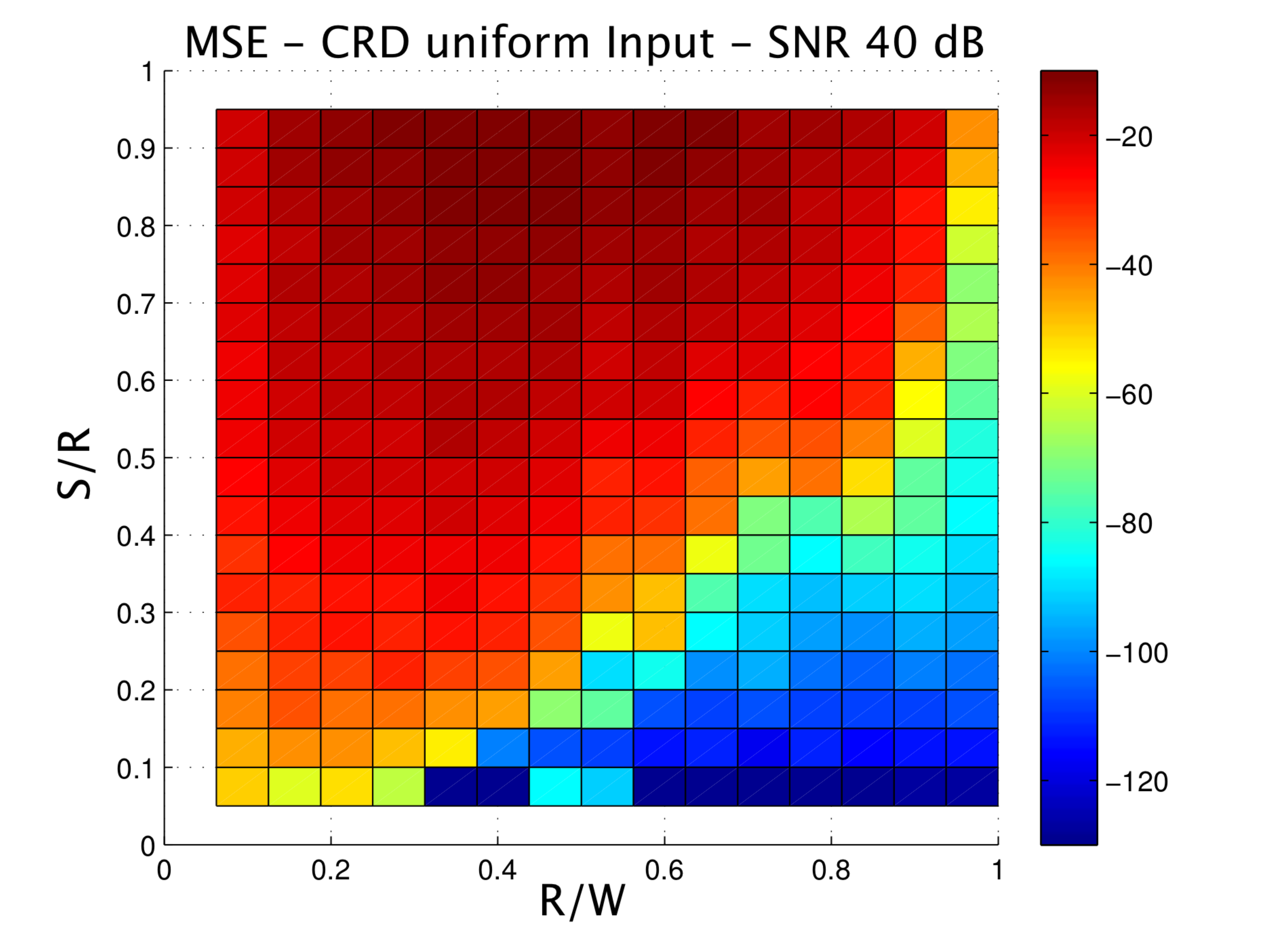}
    \label{fg:crd_MSE_uniform}
  }
  \subfigure[MSE plot for a CRD for signals with a distribution on the tones that matches the spectrum in Fig. \ref{fg:rll_spectrum}.]{
    \includegraphics[width=2.45in]{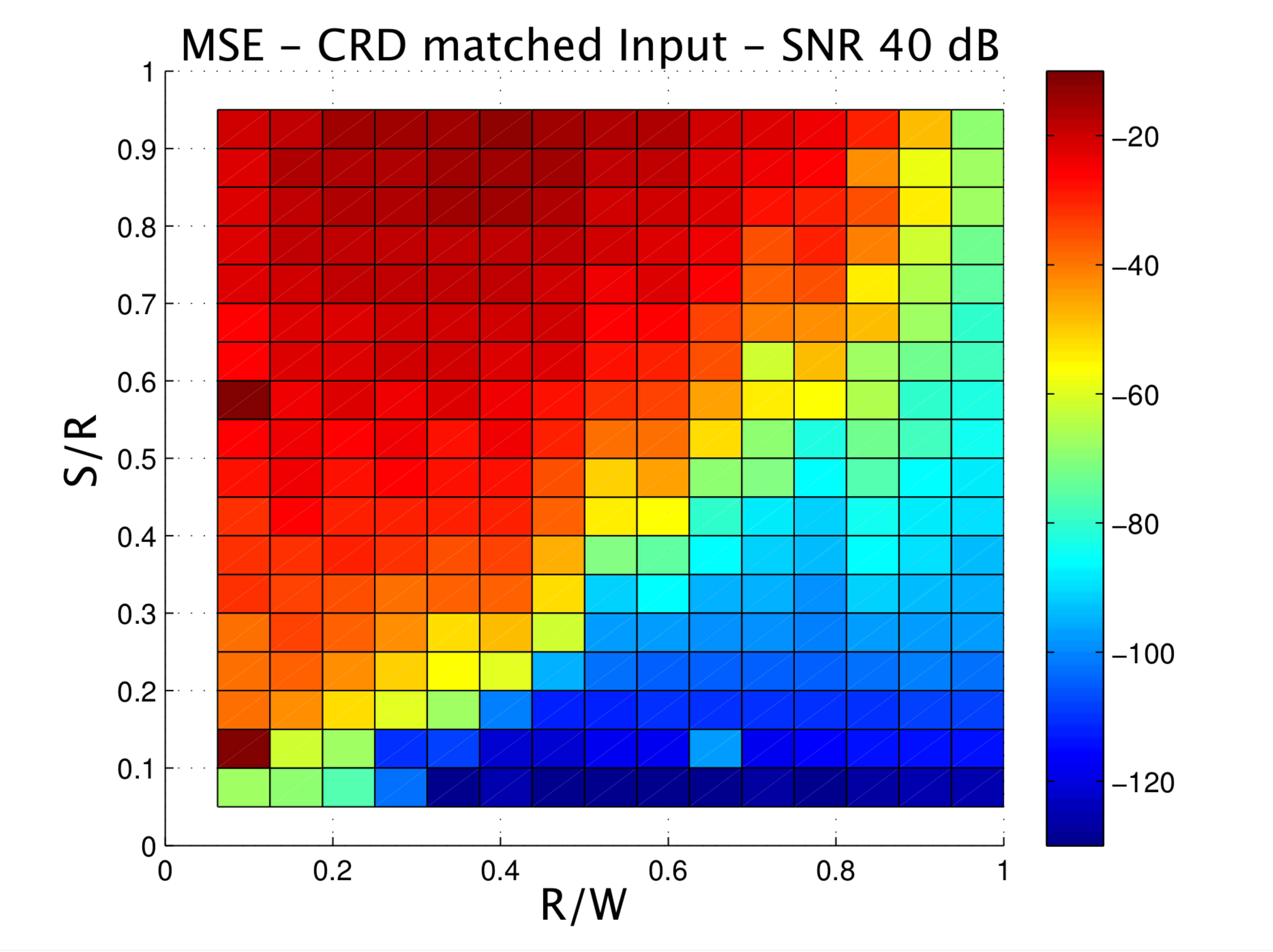}
    \label{fg:crd_MSE_matched}
  }
  \caption{Reconstruction MSE (dB) plotted as a function of $S/R$ and $R/W$.  The plots correspond to an SNR $40$ dB defined as the ratio of the power of the measurements to the noise variance.}
  \label{fg:MSE_plots}
\end{figure}

\subsection{Reconstruction in the Presence of Noise}

The phase transitions of Fig. \ref{fg:phasetransition_plots} correspond to a noiseless setting, Here, we examine the results of reconstructing input signals from noisy samples, $\mathrm{y} = \mathrm{\Phi}\mathrm{\alpha} + \sqrt{p}\cdotp\mathrm{w}$, where $\mathrm{w}$ is white Gaussian noise and $p$ determines the noise power\footnote{The model $\mathrm{y} = \mathrm{\Phi}\mathrm{(\alpha + \sqrt{p}\cdotp\mathrm{w})}$ yields similar results, but $\mathrm{w}$ as colored noise could offer interesting future work.}.  We plot the mean-squared error (MSE) of the reconstruction as a function of $S/R$ and $W/R$ and use the SpaRSA software package, which solves an $\ell_2/\ell_1$ mixed-norm optimization termed \emph{lasso} \cite{tibshirani-lasso} for noisy reconstruction purposes \cite{sparsa09}\footnote{SpaRSA is better suited for noisy reconstruction than YALL1.  For the regularization parameter, we used $1.9\sqrt{2p\log{W}}$.}; the results are shown in Fig. \ref{fg:MSE_plots}.  Similar to the noiseless case, we see a sharp transition from low MSE to high MSE.  The performance of the RD is also similar for each class of input signals while the CRD performs much better for the second class of input signals due to matching the prior to the power spectrum of the modulating sequence.

\subsection{Reconstruction of Signals with Non-Integral Frequencies}

\begin{figure}[!t] 
  \centering
  \subfigure[MSE plot for a RD for signals with a distribution on the tones that matches the spectrum in Fig. \ref{fg:rll_spectrum} and with frequency leakage.]{
    \includegraphics[width=2.8in]{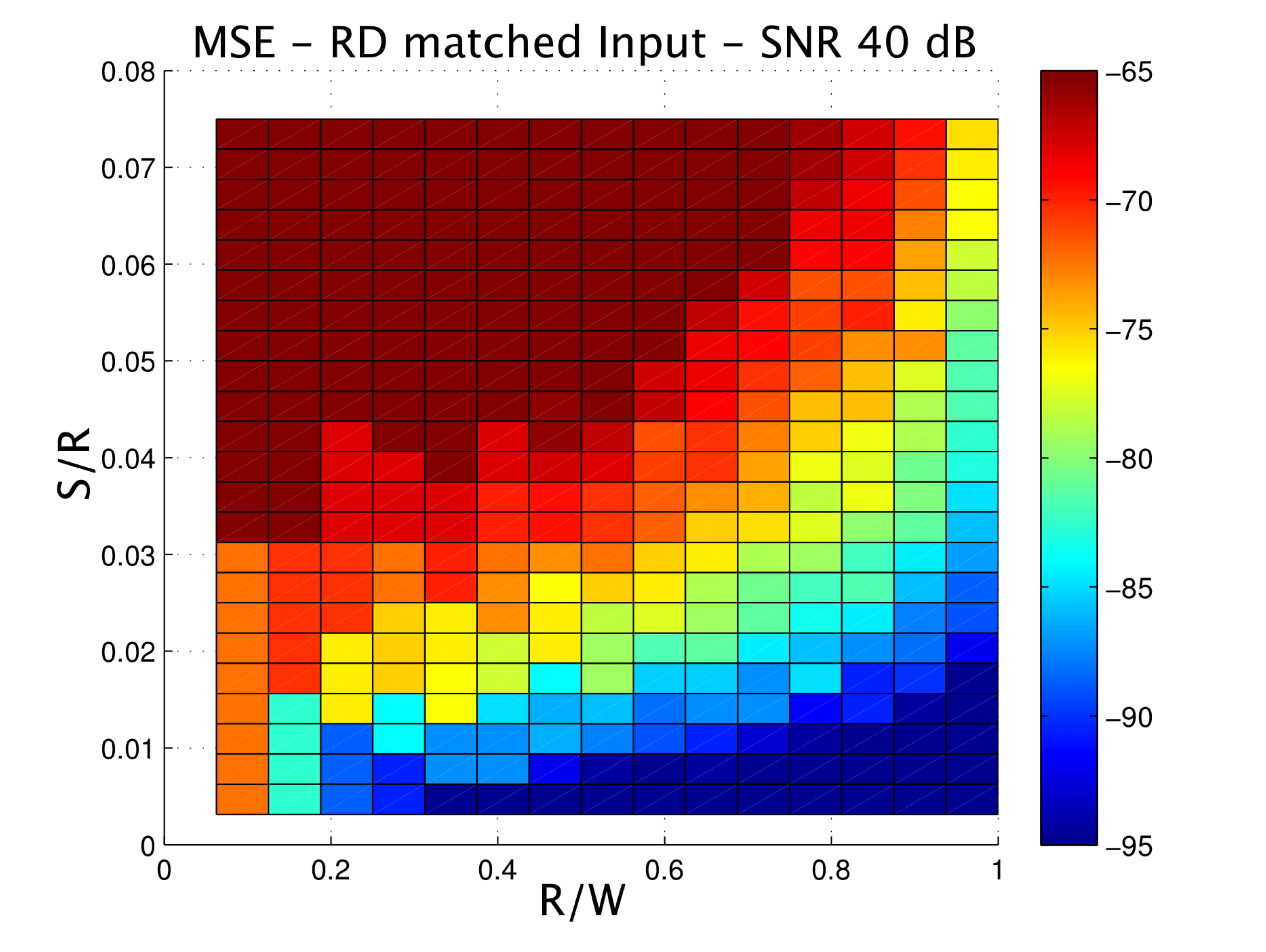}
    \label{fg:rd_MSE_matched_leakage}
  }
  \subfigure[MSE plot for a CRD for signals with a distribution on the tones that matches the spectrum in Fig. \ref{fg:rll_spectrum} and with frequency leakage.]{
    \includegraphics[width=2.8in]{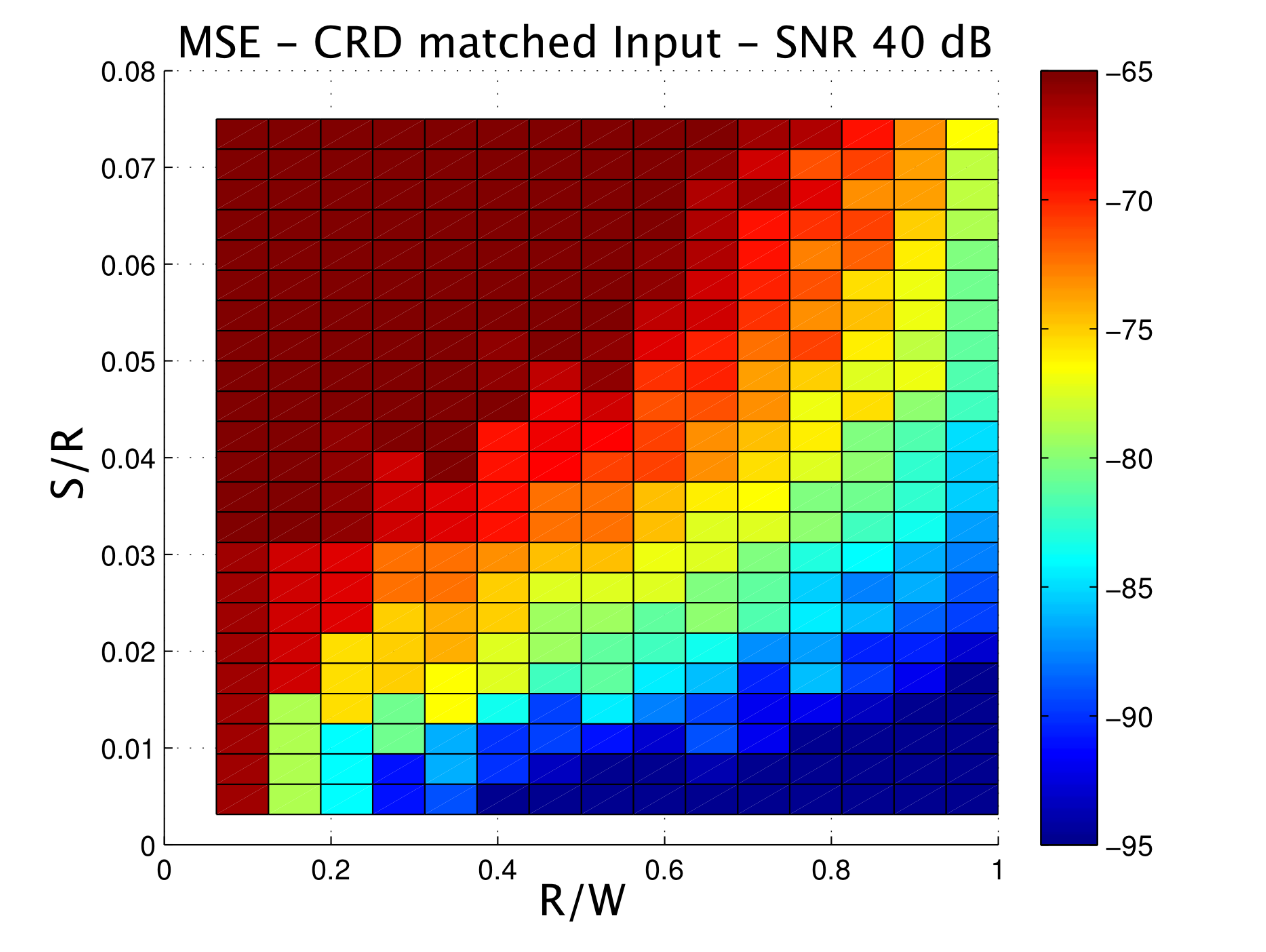}
    \label{fg:crd_MSE_matched_leakage}
  }
  \caption{Reconstruction MSE (dB) as a function of $S/R$ and $R/W$.  The plots correspond to an SNR defined as in Fig. \ref{fg:MSE_plots}.  In these experiments, non-integral frequencies are allowed and place energy at integral frequencies according to a Hamming window frequency response. }
  \label{fg:MSE_leakage_plots}
\end{figure}

The signal model \eqref{eq:sig_model} assumes only integral-frequency tones.  Real-world signals may contain non-integral frequency tones.  These non-integral tones will `leak' energy to several integral tones based on the implicit windowing operation from the finite time assumption ($t\in[0,1)$).  The windowing produces a convolution of the input tones and the window in frequency\cite{dsp-proakis} but does not invalidate the signal model \eqref{eq:sig_model}.  Rather, the result is a scaling of the sparsity factor from $S$ to $aS$, where $a \geq 1$ determines the extent of the leakage.  Fig. \ref{fg:MSE_leakage_plots} shows reconstruction results if non-integral tones in the input signal are allowed.  Tones are drawn at random from $[0,W)$ according to a distribution proportional to the spectrum in Fig. \ref{fg:rll_spectrum}.  The coefficients in the input at the integral tones are determined by a Hamming window (in the frequency domain) centered at the location of the tone.  Now, compare Fig. \ref{fg:rd_MSE_matched_leakage} with Fig. \ref{fg:rd_MSE_matched} (for the RD) and Fig. \ref{fg:crd_MSE_matched_leakage} with Fig. \ref{fg:crd_MSE_matched} (for the CRD).  Both plots look similar, but notice that  Fig. \ref{fg:MSE_leakage_plots} has $S$ scaled by a factor of 16.  This suggests that the penalty for considering leakage in \eqref{eq:sig_model} is roughly a factor of 16 in input signal sparsity.  In the worst-case, this kind of `mismatch' can seriously degrade reconstruction performance\cite{chi11}. However, in our experiments we do not often see the worst case (a tone occurring halfway between two integral tones) and hence only see a manageable decrease in performance.

\section{Conclusions}
In summary, we have proposed the use of RLL sequences in the RD because of hardware constraints on generating high-fidelity, fast-switching waveforms.  We have shown both theoretically and numerically that for a fixed switching rate, certain classes of RLL sequences offer an increase in the observable bandwidth of the system.  Specifically, we showed that an MRS works well and an RCS does not.  Insight into why each sequence succeeds or fails is found in the power spectrum of the sequence.  Further, we have argued that matching the distribution of tones in the input signal to the power spectrum of these RLL sequences improves performance, sometimes even beyond that of the RD. The most obvious future directions to take are a better theoretical understanding of knowledge-enhanced CRD and matching the modulating sequence to arbitrary distributions on the input tones.  A more thorough understanding of the hardware system and the consideration of a more complex modulating waveform (e.g., with a pulse shape other than a square) would also be interesting and useful.

\appendices
\section{Restricted Isometry Property of the CRD}\label{app:rip-proof}
To show that a CRD satisfies the RIP, we follow the proof technique of \cite{tropp10} for the RD with changes to account for correlations within $\varepsilon$ in our case.  We begin by bounding the entries of $\mathrm{\Phi_{CRD}}$.

\begin{lemma} 
\emph{[A Componentwise Bound]}
\label{lemma:comp_bound}
Let $\mathrm{\Phi_{CRD}}$ be an $R \times W$ CRD matrix, and let $\ell$ be the maximum dependence distance of the corresponding modulating sequence.  When $2 \leq p \leq 4\log W$, we have
$$\mathbb{E}^p\lVert\mathrm{\Phi_{CRD}}\rVert_{\max} \leq \sqrt{\frac{\ell\cdotp6\log W}{R}}$$
and
$$\mathbb{P}\left\{ \lVert\mathrm{\Phi_{CRD}}\rVert_{\max} > \sqrt{\frac{\ell\cdotp10\log W}{R}}\right\} \leq W^{-1}.$$
\end{lemma}

\begin{IEEEproof}
We use  the following Lemma of Tropp et al. \cite[Lemma 5]{tropp10}.
\begin{lemma} \emph{[Bounded Entries -- RD]} \label{lemma:RD-bounded}
   Let $\mathrm{\Phi_{RD}}$ be an $R\times W$ RD matrix.  When $2 \leq p \leq 4\log W$, we have
$$\mathbb{E}^p\lVert\mathrm{\Phi_{RD}}\rVert_{\max} \leq \sqrt{\frac{6\log W}{R}}$$
and
$$\mathbb{P}\left\{\lVert\mathrm{\Phi_{RD}}\rVert_{\max} > \sqrt{\frac{10\log W}{R}}\right\} \leq W^{-1}.$$
\end{lemma}

We assume that $R$ divides $W$ and $\ell$ divides $\frac{W}{R}$.  We can write each entry of $\mathrm{\Phi_{CRD}}$ as
\begin{align}
  \varphi_{r\omega} &= \sum_{j\sim r}\varepsilon_jf_{j\omega} \notag \\
  &= \sum_{(j\sim r)_0}\varepsilon_jf_{j\omega} + ... + \sum_{(j\sim r)_{\ell-1}}\varepsilon_jf_{j\omega} \notag \\
  &= \varphi_{r\omega}^{(0)} + ... + \varphi_{r\omega}^{(\ell-1)} \label{eq:split-series}
\end{align}
where $[\varepsilon_j]$ is the modulating sequence, $[f_{j\omega}] $ are the entries of the Fourier matrix $\mathrm{F}$, and $(j\sim r)_m$ denotes all $j$ such that $j\sim r$ and $(j \mod \ell) = m$.  Note that each $\varphi_{r\omega}^{(m)}$ in \eqref{eq:split-series} is a Rademacher series containing $W/R\ell$ terms, and we proceed by applying the triangle inequality to \eqref{eq:split-series}:
$$\mathbb{E}^p\varphi_{r\omega} = \mathbb{E}^p\sum_{m=0}^{\ell-1}\varphi_{r\omega}^{(m)} \leq \sum_{m=0}^{\ell-1}\mathbb{E}^p\varphi_{r\omega}^{(m)}.$$
Applying Lemma \ref{lemma:RD-bounded} to each entry in the sum, we have
$$\mathbb{E}^p ||\mathrm{\Phi_{CRD}}||_{\max} \leq \sum_{m=0}^{\ell-1}\sqrt{\frac{6\log W}{\ell R}} = \sqrt{\frac{6\ell\log W}{ R}}.$$
For the probability bound, we apply Markov's inequality.  Let $M = \lVert\mathrm{\Phi_{CRD}}\rVert_{\max}$, then
$$\mathbb{P}\left\{M > u\right\} = \mathbb{P}\left\{M^q > u^q\right\} \leq \left[\frac{\mathbb{E}^qM}{u}\right]^q$$
and choosing $u=e^{0.25}\mathbb{E}^qM$, we obtain
\begin{equation}\label{eq:maxentryTail}
  \mathbb{P}\left\{M > 2^{1.25}e^{0.25}\sqrt{\frac{\ell\log W}{R}} \right\} \leq e^{-\log W} = W^{-1}.
\end{equation}
Finally, a numerical bound yields the desired result.
\end{IEEEproof}

To complete the proof of Theorem \ref{thm:ripcord}, recall that the RIP of order $S$ with constant $\delta_S \in (0,1)$ holds if
$$|||\mathrm{\Phi_{CRD}}^*\mathrm{\Phi_{CRD}} - \mathrm{I}||| < \delta_S.$$
Using (\ref{eq:RIP-with-Delta}), we want to show that
\begin{equation}\label{eq:RIPplusDelta}
  |||\mathrm{\Phi^*_{CRD}}\mathrm{\Phi_{CRD}} - \mathbb{E}[\mathrm{\Phi^*_{CRD}}\mathrm{\Phi_{CRD}}]||| + |||\Delta||| < \delta_S.
\end{equation}
We have already bounded $|||\Delta|||$ in Section \ref{sec:general-sequences}.  We bound the first term by leveraging the results of \cite{tropp10} along with an argument similar to that used in \cite{bajwa09} for proving the RIP of Toeplitz matrices.  Before we continue, recall that the separation between two rows of $\mathrm{\Phi_{CRD}}$ required for independence between the rows is $\rho = \lceil\frac{R}{W}(\ell-1)\rceil \leq (\ell - 1)$.  In what follows, let $z_r^*$ denote the $r^{th}$ row of $\mathrm{\Phi_{RD}}$ or $\mathrm{\Phi_{CRD}}$ depending on the context.  Note that $z_rz_r^*$ is a rank one matrix and that $\mathrm{\Phi^*_{RD}}\mathrm{\Phi_{RD}} = \sum_{r=1}^{R}z_rz_r^*.$  We now need the following proposition which is a corollary to \cite[Theorems 16 and 18]{tropp10}.

\begin{proposition}\label{prop:RD-matrix}
  Let $\mathrm{\Phi_{RD}}$ be an $R\times W$ random demodulator matrix and $z_r'$ be an independent copy of $z_r$.  Define the random variable
\begin{align*}
  Z_{\mathrm{RD}} &= |||\mathrm{\Phi_{RD}}^*\mathrm{\Phi_{RD}} - \mathbb{E}\mathrm{\Phi_{RD}}^*\mathrm{\Phi_{RD}}||| \\
  &= \left|\left|\left|\sum_r\left(z_r z^*_r - \mathbb{E}z'_r{z}'^{*}_r\right)\right|\right|\right|.
\end{align*}
Then $Z_{\mathrm{RD}}$ satisfies
\begin{itemize}
  \item $\mathbb{E}Z_{\mathrm{RD}} \leq (\mathbb{E}B^2)^{1/2}\sqrt{{\mathrm{C}S\log^4W}} \leq \sqrt{\frac{\mathrm{C}S\log^5W}{R}} < \delta$, and
  \item $\mathbb{P}\{Z_{\mathrm{RD}} > \delta\} \leq 8W^{-1}$,
\end{itemize}
provided $R \geq \mathrm{C}\delta^{-2}\cdot S \log^6(W)$.  Note that
$$B = \max_{r,\omega}|\varphi_{r\omega}| \leq \sqrt{\frac{10\log W}{R}}$$
with probability exceeding $1-W^{-1}$.
\end{proposition}

To bound the first term in \eqref{eq:RIPplusDelta}, we proceed as follows
\begin{align*}
  Z_{\mathrm{CRD}} &= |||\mathrm{\Phi^*_{CRD}}\mathrm{\Phi_{CRD}} - \mathbb{E}\mathrm{\Phi^*_{CRD}}\mathrm{\Phi_{CRD}}||| \\
  &= \left|\left|\left|\sum_{r=1}^{R}z_r z_r^* - \mathbb{E}\sum_{r=1}^{R}z'_r {z}'^{*}_r\right|\right|\right| \\
  &= \left|\left|\left| \sum_{s = 1}^{\rho+1}\left(\sum_{r\in R_s}z_r z^*_r - \mathbb{E}z'_r {z}'^*_r\right)\right|\right|\right|
\end{align*}
where $R_s = \{(\rho +1)n + s\}$, $n= 0,1,...,\frac{R}{\rho+1} - 1$.  The triangle inequality tells us that
\begin{equation*}
  Z_{\mathrm{CRD}} \leq \sum_{s=1}^{\rho+1}\left|\left|\left| \sum_{r\in R_s}z_r z^*_r - \mathbb{E}z'_r{z}'^*_r \right|\right|\right| = \sum_{s=1}^{\rho+1} Z_s.
\end{equation*}

Each $Z_s$ is the norm of a sum of independent random variables, and we can apply Proposition \ref{prop:RD-matrix} to each of them.  Using Lemma \ref{lemma:comp_bound} to obtain the value of $B$ needed in Proposition \ref{prop:RD-matrix}, we get
\begin{align*}
  \mathbb{E}Z_{\mathrm{CRD}} &\leq \sum_{s=1}^{\rho+1}\mathbb{E}Z_s \leq \sum_{s=1}^{\rho+1}\sqrt{\frac{\mathrm{C}\cdotp\ell S\log^5W}{R}} \\
  &= (\rho+1)\sqrt{\frac{\mathrm{C}\cdotp\ell S\log^5W}{R}}.
\end{align*}
We require that $\mathbb{E}Z_{\mathrm{CRD}} < \delta'$ for $\delta' \in (0,1)$ which is achieved as long as
$$R\geq \mathrm{C}\ell(\rho+1)^2(\delta')^{-2}S\log^5W.$$
We can similarly appeal to the probability bound in Proposition \ref{prop:RD-matrix} to obtain
$$\mathbb{P}\{Z_s > \delta'/(\rho+1)\} \leq 8W^{-1}$$
if $R\geq\mathrm{C}\ell(\rho+1)^2(\delta')^{-2}S\log^6W$.  Returning to \eqref{eq:RIPplusDelta}, we have
$$|||\mathrm{\Phi^*_{CRD}}\mathrm{\Phi_{CRD}} - \mathrm{I}||| < \delta$$
if $\delta' < (\delta - |||\Delta|||)$, and the RIP of order $S$ is satisfied with constant $\delta_S \leq \delta$ completing the proof of Theorem \ref{thm:ripcord}.

\section{Recovery under the Random Signal Model}\label{app:random-signal-model}
To prove Theorem \ref{thm:rand-sig-coherence}, we must bound the coherence and column norms of the matrix $\mathrm{\Phi_{CRD}}$.  To bound the coherence, we bound the maximum absolute entry of $\mathrm{X}$ \eqref{eq:matrixX}:
$$\max_{\alpha,\omega}|x_{\alpha,\omega}| = \max_{\alpha,\omega} \left|\sum_{j\neq k}\varepsilon_j\varepsilon_k\eta_{jk}f^*_{j\alpha}f_{k\omega}\right|.$$
If the sequence $\varepsilon$ is not independent, but has maximum dependence distance $\ell$, then we need to break the sum up into smaller sums.  Define the sets $J_a = \{n\ell + a\}$, $0 \leq a \leq \ell-1$, $0 \leq n \leq \frac{W}{\ell}-1$ and $K_j = \{j-(\ell-1), ..., j+(\ell-1)\}$.  We now apply the triangle inequality twice to $\left|x_{\alpha\omega}\right|$:
\begin{align*}
  & |x_{\alpha\omega}| = \left|\sum_{j\neq k}\varepsilon_j\varepsilon_k\eta_{jk}f^*_{j\alpha}f_{k\omega}\right| \\
  &= \left| \sum_{j}\left(\sum_{\substack{k\in K_j \\ k\neq j}}\varepsilon_j\varepsilon_k\eta_{jk} f^*_{j\alpha}f_{k\omega}\right) + \left(\sum_{k\not\in K_j}\varepsilon_j\varepsilon_k\eta_{jk} f^*_{j\alpha}f_{k\omega}\right) \right| \\
  &\leq \left| \sum_{j}\sum_{\substack{k\in K_j \\ k\neq j}}\varepsilon_j\varepsilon_k\eta_{jk} f^*_{j\alpha}f_{k\omega}\right| + \left|\sum_{a=0}^{\ell-1}\left[\sum_{\substack{j\in J_a, \\ k\not\in K_j}}\varepsilon_j\varepsilon_k\eta_{jk} f^*_{j\alpha}f_{k\omega}\right]\right| \\
  &\leq \left| \sum_{j}\sum_{\substack{k\in K_j \\ k\neq j}}\varepsilon_j\varepsilon_k\eta_{jk} f^*_{j\alpha}f_{k\omega}\right| + \sum_{a=0}^{\ell-1}\left|\sum_{\substack{j\in J_a, \\ k\not\in K_j}}\varepsilon_j\varepsilon_k\eta_{jk} f^*_{j\alpha}f_{k\omega}\right| \\
  &= E + \sum_{a=0}^{\ell-1}M_a.
\end{align*}

Each $M_a$ is a second-order Rademacher chaos because of the indices of summation, $J_a$ and $K_j$, and we need the following to deal with such a sum.

\begin{proposition}\emph{\cite[Lemma 6]{tropp10}}
  Suppose that $R\geq 2\log W$.  Let $[\varepsilon_j]$ be an independent modulating sequence and define $x_{\alpha\omega} = \sum_{j\neq k}\varepsilon_j\varepsilon_k\eta_{jk}f^*_{j\alpha}f_{k\omega}$ and $\mathrm{X} = [x_{\alpha\omega}]$.  Then
$$\mathbb{E}^p[||\mathrm{X}||_{\max}] \leq 8\mathrm{C}\sqrt{\frac{\log W}{R}}$$
and
$$\mathbb{P}\left\{||\mathrm{X}||_{\max} > \mathrm{C}\sqrt{\frac{\log W}{R}} \right\} \leq W^{-1}.$$

\end{proposition}
Applying this proposition to each $M_a$, we get
$$\mathbb{E}^pM_a \leq 8\mathrm{C}\sqrt{\frac{\log W}{R}} \Rightarrow \mathbb{E}^p\left[\sum_{a=0}^{\ell-1}M_a\right] \leq 8\mathrm{C}\ell\sqrt{\frac{\log W}{R}}.$$
It follows from Markov's inequality that
$$\mathbb{P}\left\{\sum_{a=0}^{\ell-1}M_a > \mathrm{C}\ell\sqrt{\frac{\log W}{R}} \right\} \leq W^{-1}.$$

Now we are left to deal with $E$.  Whenever $W/R \geq \ell$ we can drop $\eta_{jk}$ because $\eta_{jk} = 1$ over the index of summation.  We can then rewrite $E$ in this case as follows:
\begin{align*}
  E &= \left| \sum_{j}\sum_{k\in K_j, k\neq j}\varepsilon_j\varepsilon_k\eta_{jk} f^*_{j\alpha}f_{k\omega}\right| \\
  &= \left| \sum_{j}\varepsilon_j f^*_{j\alpha} \left(\sum_{k\in K_j, k\neq j}\varepsilon_k f_{k\omega}\right)\right| \\
  &= \left| \sum_{j}\varepsilon_j f^*_{j\alpha} \left|E^{(j)}_2\right|\exp\left(\imath \cdotp \text{phase}\left(E^{(j)}_2\right)\right)\right| \\
  & = \left| \sum_{j}\varepsilon_j f'_{j\alpha} \left|E^{(j)}_2\right|\right|
\end{align*}
where
$$E^{(j)}_2 = \sum_{k\in K_j, k\neq j}\varepsilon_k f_{k\omega},$$
$\text{phase}(\cdot)$ is the phase angle of the complex argument, and $f'_{j\alpha} = f^*_{j\alpha}\cdotp\exp\left(\imath \cdotp \text{phase}\left(E^{(j)}_2\right)\right)$.  In short order, we will bound $|E^{(j)}_2| \leq t_2$ $\forall j$ with high probability so that $E$ can be bounded as
$$E \leq \left| \sum_{j}\varepsilon_j f'_{j\alpha} \right| \cdotp t_2 = E_1 \cdotp t_2$$
with high probability.  To bound $E_1$ and to find $t_2$, we turn to a result to bound the norm of a random series generated from a Markov chain.

\begin{proposition}\emph{\cite[Corollary 4]{samson00}}
  Let $\varepsilon = [\varepsilon_{j}]$ be a sequence of random variables generated from a Markov chain with $\varepsilon_j \in \{+1,-1\}$ equally likely.  Let the matrix $\Gamma$ be the matrix defined in Section \ref{sec:main-results}.  Let $b_i$ for $1\leq i\leq n$ be arbitrary complex numbers and let $f = \left|\sum_{i=1}^{n}\varepsilon_i b_i \right|$.  For every $t\geq 0$,
$$\mathbb{P}\left(|f-\mathbb{E}[f]| \geq t\right) \leq \exp\left(-\frac{t^2}{8\sigma^2||\Gamma||^2}\right)$$
where
$$\sigma^2 = \sum_{i=1}^{n} |b_i|^2.$$

\end{proposition}

We apply this proposition to both $E_1$ and $|E^{(j)}_2|$ with
$$t_1 = \sqrt{\log W \cdotp 8\sigma_1^2||\Gamma||^2}$$
and
$$t_2 = \sqrt{\log W \cdotp 16\sigma_2^2||\Gamma||^2}$$
respectively.  As a result,
$$\mathbb{P}\left(E_1 \geq t_1\right) \leq \exp(-\log W) = W^{-1}$$
and
$$\mathbb{P}\left(|E^{(j)}_2| \geq t_2\right) \leq W^{-2}$$
$\forall j$.  Finally, we have that $E \leq t_1\cdotp t_2$ except with probability $2W^{-1}$.  To finish the calculation, note that
$$\sigma_1^2 = \sum_{j=0}^{W-1}|f_{j\alpha}^*|^2 = 1$$
and
$$\sigma_2^2 = \sum_{k\in K_{\ell}, k\neq \ell}|f_{k\omega}|^2 = 2(\ell-1)/W.$$
Hence,
\begin{align*}
  t_1\cdotp t_2 &= \sqrt{\log W \cdotp 8\sigma_1^2||\Gamma||^2} \cdotp \sqrt{\log W \cdotp 16\sigma_2^2||\Gamma||^2} \\
  &= \log W \cdotp 8\sqrt{2}||\Gamma||^2 \sqrt{\sigma_1^2\sigma_2^2} \\
  &= \frac{\log W}{\sqrt{W}} 16\sqrt{\ell-1}||\Gamma||^2.
\end{align*}

Finally, we have the following for the matrix $\mathrm{X}$:
\begin{equation*}
  \mathbb{P}\left(||\mathrm{X}||_{\max} \geq \mathrm{C}\ell\sqrt{\frac{\log W}{R}} + t_1\cdotp t_2 \right) \leq 3W^{-1}.
\end{equation*}
Note that $\lim_{W\to\infty} (\log W / \sqrt{W}) = 0$, so we can make the second term as small as we like by requiring a large enough $W$.  This leads us to the following statements about the \emph{coherence}, $\mu = \max_{\alpha\neq\omega}|\langle\phi_{\alpha},\phi_{\omega}\rangle|$, and \emph{column norms} of a CRD matrix:

\begin{lemma}\label{thm:crd_coherence}
  \emph{[Coherence]} Suppose that $R \geq 2\log W$.  An $R\times W$ CRD matrix satisfies
$$\mathbb{P}\left(\mu \geq \mathrm{C}\ell\sqrt{\frac{\log W}{R}} + \frac{\log W}{\sqrt{W}} 16\sqrt{\ell-1}||\Gamma||^2 \right) \leq 3W^{-1}.$$
\end{lemma}

\begin{lemma}\label{thm:crd_colnorm}
  \emph{[Column Norms]}  Suppose the sampling rate satisfies
$$R \geq 4\cdotp\mathrm{C}\ell^2\delta^{-2}\log W$$
and that $W$ is large enough so that
$$\frac{\log(W)}{\sqrt{W}} \leq \frac{\delta}{32\sqrt{(\ell-1)}||\Gamma||^2}.$$
Then, an $R\times W$ CRD matrix satisfies
$$\mathbb{P}\left\{\max_{\omega}\left| ||\phi_{\omega}||_2^2 -1 \right| \geq \delta \right\} \leq 3W^{-1}.$$
\end{lemma}

To prove recovery results, we finally use the following theorem.

\begin{theorem}\label{thm:tropp_randsig}\emph{\cite[Corollary 15]{tropp10}}
  Suppose that the sampling rate satisfies
$$R \geq \mathrm{C}[S\log W + \log^3 W].$$
Draw an $R\times W$ RD matrix such that
$$\mathbb{P}\left\{\mu \geq \mathrm{C}\sqrt{\frac{\log W}{R}}\right\} \leq W^{-1}$$
and
$$\mathbb{P}\left\{\max_{\omega}\left|||\phi_{\omega}||^2_2 - 1 \right| \geq \delta\right\} \leq W^{-1}.$$
Let $\mathrm{s}$ be an $S$-sparse vector drawn according to the random signal model in Section \ref{sec:background}.  The solution $\hat{\mathrm{s}}$ to the convex program \eqref{eq:convex_program} satisfies $\hat{\mathrm{s}} = \mathrm{s}$ except with probability $8W^{-1}$.

\end{theorem}

Theorem \ref{thm:rand-sig-coherence} is the result of applying Lemmata \ref{thm:crd_coherence} and \ref{thm:crd_colnorm} to Theorem \ref{thm:tropp_randsig}.  The increased requirement on $R$ and the additional requirement on $W$ is needed to ensure the coherence and column norms are satisfactory to ensure recovery.  Additionally, the probability of recovery failing increases slightly to $12W^{-1}$.

\section{Uncorrelated implies Independence for Identically Distributed bipolar sequences}\label{app:corr-vs-dep}
Here we briefly show that if two entries in the modulating sequence are uncorrelated then they are independent for the sequences that arise in this paper.  The sequences, denoted by $[\varepsilon_j]$ for $j=1,...,W$, that we are concerned with have two defining characteristics:
($i$) $\varepsilon_j \in \{+1,-1\}$ and ($ii$) $\mathbb{P}\{\varepsilon_j = +1\} = \mathbb{P}\{\varepsilon_j = -1\} = 1/2$.
The autocorrelation in this case can be expressed as:
$$\mathbb{E}[\varepsilon_j\varepsilon_{j+k}] = \mathbb{P}\{\varepsilon_j = \varepsilon_{j+k}\} - \mathbb{P}\{\varepsilon_j \neq \varepsilon_{j+k}\}.$$
Now, given the maximum dependence distance $\ell$ we have $\mathbb{P}\{\varepsilon_j = \varepsilon_{j+k}\} = \mathbb{P}\{\varepsilon_j \neq \varepsilon_{j+k}\}$ for $|k| \geq \ell$ which implies that
$$\mathbb{P}\{\varepsilon_{j+k} = +1 | \varepsilon_j = +1\} = \mathbb{P}\{\varepsilon_{j+k} = +1 | \varepsilon_j = -1\}$$
in this case.  Characteristic (ii) also tells us that
$$\mathbb{P}\{\varepsilon_{j+k} = +1 | \varepsilon_j = +1\} + \mathbb{P}\{\varepsilon_{j+k} = +1 | \varepsilon_j = -1\} = 1,$$
meaning we must have that
$$\mathbb{P}\{\varepsilon_{j+k} = +1 | \varepsilon_j = +1\} = \mathbb{P}\{\varepsilon_{j+k} = +1 | \varepsilon_j = -1\} = 1/2.$$
The same argument applies to $\varepsilon_{j+k} = -1$, and the condition for independence results:
$$\mathbb{P}\{\varepsilon_{j+k} = a| \varepsilon_j = b\} = \mathbb{P}\{\varepsilon_{j+k} = a\}$$
for $a,b \in \{+1,-1\}$.

\bibliographystyle{IEEEtran}
\bibliography{IEEEabrv,andrew}

\begin{thebibliography}{10}
\providecommand{\url}[1]{#1}
\csname url@samestyle\endcsname
\providecommand{\newblock}{\relax}
\providecommand{\bibinfo}[2]{#2}
\providecommand{\BIBentrySTDinterwordspacing}{\spaceskip=0pt\relax}
\providecommand{\BIBentryALTinterwordstretchfactor}{4}
\providecommand{\BIBentryALTinterwordspacing}{\spaceskip=\fontdimen2\font plus
\BIBentryALTinterwordstretchfactor\fontdimen3\font minus
  \fontdimen4\font\relax}
\providecommand{\BIBforeignlanguage}[2]{{%
\expandafter\ifx\csname l@#1\endcsname\relax
\typeout{** WARNING: IEEEtran.bst: No hyphenation pattern has been}%
\typeout{** loaded for the language `#1'. Using the pattern for}%
\typeout{** the default language instead.}%
\else
\language=\csname l@#1\endcsname
\fi
#2}}
\providecommand{\BIBdecl}{\relax}
\BIBdecl

\bibitem{le05}
B.~Le, T.~W. Rondeau, J.~H. Reed, and C.~W. Bostian, ``Analog-to-digital
  converters: {A} review of the past, present, and future,'' \emph{{IEEE} Sig.
  Proc. Mag.}, pp. 69--77, Nov. 2005.

\bibitem{walden08}
R.~H. Walden, ``Analog-to-digital converters and associated {IC}
  technologies,'' in \emph{Proc. {IEEE} CSICS}, Oct. 2008, pp. 1--2.

\bibitem{murmannADC}
\BIBentryALTinterwordspacing
B.~Murmann. {ADC} performance survey 1997-2012. [Online]. Available:
  \url{http://www.stanford.edu/~murmann/adcsurvey.html}
\BIBentrySTDinterwordspacing

\bibitem{yoo-candes12}
M.~Wakin, S.~Becker, E.~Nakamura, M.~Grant, E.~Sovero, D.~Ching, J.~Yoo,
  J.~Romberg, A.~Emami-Neyestanak, and E.~Candes, ``A non-uniform sampler for
  wideband spectrally-sparse environments,'' \emph{Submitted to IEEE JETCAS},
  2012.

\bibitem{candes-romberg-tao06}
E.~Cand\`{e}s, J.~Romberg, and T.~Tao, ``Robust uncertainty principles: Exact
  signal reconstruction from highly incomplete frequency information,''
  \emph{IEEE Trans. Inform. Theory}, vol.~52, no.~2, pp. 489--509, Feb 2006.

\bibitem{tropp10}
J.~Tropp, J.~Laska, M.~Duarte, J.~Romberg, and R.~Baraniuk, ``Beyond {N}yquist:
  Efficient sampling of sparse bandlimited signals,'' \emph{{IEEE} Trans.
  Inform. Theory}, vol.~56, no.~1, pp. 520--544, Jan. 2010.

\bibitem{martin2012}
T.~Carusone, D.~Johns, and K.~Martin, \emph{Analog Integrated Circuit
  Design}.\hskip 1em plus 0.5em minus 0.4em\relax John Wiley \& Sons, 2012.

\bibitem{tang70}
D.~Tang and L.~Bahl, ``Block codes for a class of constrained noiseless
  channels,'' \emph{Inform. Cont.}, pp. 436--461, Dec 1970.

\bibitem{immink98}
K.~Immink, P.~Siegel, and J.~Wolf, ``Codes for digital recorders,''
  \emph{{IEEE} Trans. Inform. Theory}, pp. 2260--2299, Oct 1998.

\bibitem{harms11-icassp}
A.~Harms, W.~U. Bajwa, and R.~Calderbank, ``Beating {N}yquist through
  correlations: A constrained random demodulator for sampling of sparse
  bandlimited signals,'' in \emph{Proc. {IEEE} {ICASSP}}, 2011.

\bibitem{harms-camsap11}
------, ``Faster than {N}yquist, slower than {T}ropp,'' in \emph{Proc. {IEEE}
  {CAMSAP}}, 2011.

\bibitem{donoho09}
D.~Donoho and J.~Tanner, ``Observed universality of phase transitions in
  high-dimensional geometry, with implications for modern data analysis and
  signal processing,'' \emph{Phil. Trans. A Math. Phys. Eng. Sci.}, pp.
  4273--93, Nov 2009.

\bibitem{setti10}
J.~Ranieri, R.~Rovatti, and G.~Setti, ``Compressive sensing of localized
  signals: Application to analog-to-information conversion,'' in \emph{Proc. of
  {IEEE} ISCAS}, May 2010, pp. 3513--3516.

\bibitem{setti11}
M.~Mangia, R.~Rovatti, and G.~Setti, ``Analog-to-information conversion of
  sparse and non-white signals: Statistical design of sensing waveforms,'' in
  \emph{Proc. of {IEEE} ISCAS}, May 2011, pp. 2129--2132.

\bibitem{rife74}
D.~C. Rife and R.~R. Boorstyn, ``Single-tone parameter estimation from
  discrete-time observations,'' \emph{{IEEE} Trans. Inform. Theory}, vol.~20,
  no.~5, pp. 591--598, Sep. 1974.

\bibitem{rife76}
------, ``Multiple tone parameter estimation from discrete-time observations,''
  \emph{Bell Syst. Tech. J.}, vol.~55, pp. 1389--1410, Nov 1976.

\bibitem{duarte11}
M.~Duarte and Y.~Eldar, ``Structured compressed sensing: {F}rom theory to
  applications,'' \emph{{IEEE} Trans. Sig. Proc.}, vol.~59, no.~9, pp.
  4053--4085, Sep 2011.

\bibitem{applebaum08}
L.~Applebaum, S.~Howard, S.~Searle, and R.~Calderbank, ``Chirp sensing codes:
  Deterministic compressed sensing measurements for fast recovery,''
  \emph{Appl. Comp. Harmonic Anal.}, pp. 283--290, Sep. 2008.

\bibitem{eldar09}
M.~Mishali, Y.~Eldar, O.~Dounaevsky, and E.~Shoshan, ``Xampling: Analog to
  digital at sub-{N}yquist rates,'' \emph{{IET} J. Circ., Dev., and Sys.},
  vol.~5, no.~1, pp. 8--20, Jan 2011.

\bibitem{vetterli02}
M.~Vetterli, P.~Marziliano, and T.~Blu, ``Sampling signals with finite rate of
  innovation,'' \emph{{IEEE} Trans. Sig. Proc.}, vol.~50, no.~6, pp.
  1417--1428, June 2002.

\bibitem{unser2000}
M.~Unser, ``Sampling -- 50 years after {S}hannon,'' \emph{Proc. {IEEE}},
  vol.~88, no.~4, pp. 569--587, Apr 2000.

\bibitem{blu-unser-1999}
T.~Blu and M.~Unser, ``Quantitative {F}ourier analysis of approximation
  techniques: Part {I}--interpolators and projectors,'' \emph{{IEEE} Trans.
  Sig. Proc.}, vol.~47, no.~10, pp. 2783--2795, Oct 1999.

\bibitem{blu-unser-part2-1999}
------, ``Quantitative {F}ourier analysis of approximation techniques: Part
  {II}--wavelets,'' \emph{{IEEE} Trans. Sig. Proc.}, vol.~47, no.~10, pp.
  2796--2806, Oct 1999.

\bibitem{nichols11}
J.~M. Nichols and F.~Bucholz, ``Beating {N}yquist with light: {A} compressively
  sampled photonic link,'' \emph{Opt. Express}, vol.~19, pp. 7339--7348, 2011.

\bibitem{jyoo-RMPI}
J.~Yoo, S.~Becker, M.~Monge, M.~Loh, E.~Cand\`{e}s, and A.~Emami-Neyestanak,
  ``Design and implementation of a fully integrated compressed-sensing signal
  acquisition system,'' in \emph{Proc. {IEEE} {ICASSP}}, 2012.

\bibitem{lexa11}
M.~A. Lexa, M.~E. Davies, and J.~S. Thompson, ``Reconciling compressive
  sampling systems for spectrally-sparse continuous-time signals,''
  \emph{{arXiv}:1101.4100}, May 2011.

\bibitem{tropp08-condofrandict}
J.~A. Tropp, ``On the conditioning of random subdictionaries,'' \emph{Appl.
  Comp. Harmonic Anal.}, vol.~25, pp. 1--24, 2008.

\bibitem{candes05}
E.~J. Candes and T.~Tao, ``Decoding by linear programming,'' \emph{IEEE Trans.
  Inform. Theory}, pp. 4203--4215, Dec 2005.

\bibitem{candes-StableRecovery}
E.~J. Cand\`{e}s, J.~Romberg, and T.~Tao, ``Stable signal recovery from
  incomplete and inaccurate measurements,'' \emph{Comm. Pure Appl. Math.},
  vol.~59, pp. 1207--1223, 2005.

\bibitem{samson00}
P.-M. Samson, ``Concentration of measure inequalities for {M}arkov chains and
  $\phi$-mixing processes,'' \emph{Annals Prob.}, vol.~28, no.~1, pp. 416--461,
  2000.

\bibitem{bilardi83}
G.~Bilardi, R.~Padovani, and G.~Pierbon, ``Spectral analysis of functions of
  {M}arkov chains with applications,'' \emph{IEEE Trans. Comm.}, pp. 853--861,
  Jul. 1983.

\bibitem{hornandjohnson}
R.~A. Horn and C.~R. Johnson, \emph{Matrix Analysis}.\hskip 1em plus 0.5em
  minus 0.4em\relax Cambridge University Press, 1985.

\bibitem{yang-zhang10}
J.~Yang and Y.~Zhang, ``Alternating direction algorithms for {L1}-problems in
  compressive sensing,'' \emph{{SIAM} J. Sci. Comp.}, vol.~33, no. 1-2, pp.
  250--278, 2011.

\bibitem{baraniuk10}
R.~G. Baraniuk, V.~Cevhar, M.~F. Duarte, and C.~Hegde, ``Model-based
  compressive sensing,'' \emph{{IEEE} Trans. Inform. Theory}, vol.~56, no.~4,
  pp. 1982--20\,001, Apr 2010.

\bibitem{tibshirani-lasso}
R.~Tibshirani, ``Regression shrinkage and selection via the lasso,'' \emph{J.
  R. Statist. Soc. B}, vol.~58, no.~1, pp. 267--288, 1996.

\bibitem{sparsa09}
S.~Wright, R.~Nowak, and M.~Figueiredo, ``Sparse reconstruction by separable
  approximation,'' \emph{{IEEE} Trans. Sig. Proc.}, vol.~57, no.~7, pp.
  2479--2493, July 2009.

\bibitem{dsp-proakis}
J.~Proakis and D.~Manolakis, \emph{Digital Signal Processing}, 4th~ed.\hskip
  1em plus 0.5em minus 0.4em\relax Prentice Hall, 2006.

\bibitem{chi11}
Y.~Chi, L.~Scharf, A.~Pezeshki, and A.~Calderbank, ``Sensitivity to basis
  mismatch in compressed sensing,'' \emph{{IEEE} Trans. Sig. Proc.}, vol.~59,
  no.~5, pp. 2182--2195, May 2011.

\bibitem{bajwa09}
W.~U. Bajwa, ``New information processing theory and methods for exploiting
  sparsity in wireless systems,'' Ph.D. dissertation, University of
  Wisconsin-Madison, 2009.

\end{thebibliography}

\end{document}